\documentclass{aa}  

\usepackage{graphicx}

\usepackage{txfonts}
\usepackage{soul}
\usepackage{nicefrac}

\usepackage{hyperref}
\hypersetup{
    colorlinks=true,
    citecolor=blue,
    urlcolor=blue
    }
    
\usepackage{multirow}
\usepackage{threeparttable, tablefootnote}
\usepackage{placeins}
\usepackage{subcaption}

\graphicspath{{images/}}

\newcommand{\Msun}{\rm M_{\odot}}
\newcommand{\MsunPc}{\rm M_{\odot}~pc^{-2}}

\begin{document} 
   \title{Bar evolution in edge-on galaxies: A demographic study of boxy/peanut bulges}
   
   \titlerunning{A demographic study of boxy/peanut bulges} 
   \authorrunning{Samanta et al. 2025}

   \author{Atul A. Samanta\inst{1}\fnmsep\thanks{satulashutosh@gmail.com}
          \and
          Ankit Kumar\inst{2}\fnmsep\thanks{ankit4physics@gmail.com}
          \and
          Mousumi Das\inst{3}
          \and
          M. Celeste Artale\inst{2}
          }

   \institute{Indian Institute of Science Education and Research, Bhopal, 462066, India
         \and
         Universidad Andrés Bello, Facultad de Ciencias Exactas, Departamento de Física y Astronomía, Instituto de Astrofísica, Fernández Concha 700, Las Condes, Santiago RM, Chile
         \and
         Indian Institute of Astrophysics, Bangalore 560034, India
             }

  \abstract
   {Boxy/peanut and X-shaped (BP/X) bulges are prominent features in edge-on disk galaxies and are believed to be vertically thickened bars. Despite their relevance in bar evolution, a statistically robust census of these structures in large surveys has been lacking.}
   {We aim to provide the largest catalog of BP/X structures in edge-on galaxies to date, and to investigate their properties and role in shaping galaxy scaling relations.}
   {We selected a sample of 6684 edge-on galaxies from SDSS DR8 using Galaxy Zoo classifications, requiring a high edge-on probability ($>0.9$) and a minimum of 10 independent votes. Two-dimensional image decomposition is performed using \textsc{GALFIT} to obtain structural parameters. Residual images are visually inspected to classify BP/X features into four categories: strong both-sided, both-sided, one-sided, and control (no BP/X). We also estimate stellar mass, distance, and physical size for each galaxy.}
   {Out of 6653 classified galaxies, we identified 1673 ($\sim$25\%) with both-sided BP/X features—504 ($\sim 8\%$) strong and 1169 ($\sim 17\%$) weak—as well as 1112 ($\sim 17\%$) one-sided structures, making up a total of 2785 BP/X-hosting galaxies ($\sim 42\%$). One-sided structures, likely signatures of ongoing buckling, are more frequent than strong both-sided bulges across all stellar masses. The fraction of BP/X bulges increases with stellar surface mass density, indicating a connection with bar formation in dense disks. We also find that galaxies with strong BP/X bulges contribute to increased scatter in the stellar mass–size and stellar mass–surface density relations, particularly at higher masses.}
   {}

   \keywords{Galaxies: formation -- Galaxies: evolution -- Galaxies: bulges -- Galaxies: structure -- Galaxies: photometry -- Galaxies: statistics}

   \maketitle
%

\section{Introduction}

Spiral galaxies with elongated bar-like structures in their center are commonly known as barred galaxies and comprise nearly 30-70\% of all spiral galaxies \citep{aguerri2009population, masters2011galaxy, diaz2016characterization}. Bars play a major role in the evolution of spiral galaxies \citep{Marchuk.etal.2022}. They are thought to funnel a large amount of gas into the nuclear region \citep{Shlosman.etal.1989, Boone.etal.2007}. This gas helps in the formation of new stars in the central and bar regions \citep{Diaz-Garcia.etal.2020, Fraser-McKelvie.etal.2020}. The accretion of gas into the nuclear region also feeds the central supermassive black holes (SMBHs) that help the growth of SMBHs. Sometimes the excess gas accretion onto SMBH ignites the nuclear activity in the galaxy \citep{Alonso.etal.2013, Galloway.etal.2015, Silva-Lima.etal.2022,10.1093/mnras/stae1620, Kataria.Vivek.2024}. 

During the growth phase of bars, nuclear stellar disc (NSD) is formed due to intense star formation in the central sub-kpc region \citep{friedli1993secular, heller1994fueling, friedli1995secular, martin1997star, wozniak2007distribution, kim2011nuclear, cole2014formation, seo2019effects, baba2020age, sormani2022stellar}. The bar suppresses star formation through out the remainder of its extent
\citep{martin1997star, spinoso2016bar, khoperskov2018bar, donohoe2019redistribution}. Therefore, the NSD has younger stars than the age of the bar \citep{baba2020age}. Using their numerical simulations, \cite{baba2020age} showed that the oldest stellar population in the NSD can be used for dating bar formation time \citep[also see][]{de_Sa-Freitas.etal.2023}.

Bars drive the secular evolution of spiral galaxies by redistributing matter and angular momentum \citep{athanassoula2003determines, Minchev.etal.2011}. Despite numerous studies on bar formation and evolution, their origin remains a topic of debate \citep[see][for a review]{Sellwood2014}. The bar formation mechanisms involve global unstable modes of rotating disk \citep{Kalnajs.etal.1972, Ostriker.Peebles.1973}, swing amplification of leading small perturbations \citep{Toomre.etal.1981}, slow trapping of stellar orbits due to a weak non-axisymmetric perturbation \citep{lynden1979mechanism}, mutual gravitational interaction of precessing orbits \citep{polyachenko2003unified}, and tidal interactions between galaxies \citep{Miwa.Noguchi.1998, Lang.etal.2014, Lokas.2018}.

Studies suggest that disc evolution of the galaxy also includes the development of bars in length and thickness \citep{athanassoula2003determines, athanassoula2005nature}. The vertical thickening of bars in galaxies is mainly studied using numerical simulations \citep{combes1990box, raha1991dynamical, bureau2005bar, Martinez-Valpuesta.etal.2006, Kumar.etal.2021, kumar2022effect, Kumar2023, Ghosh.etal.2024}. Among all bar thickening mechanisms bar buckling is the most violent mechanism, in which the galactic bar bends out of the galactic plane \citep{combes1990box, raha1991dynamical, lokas2019anatomy, kumar2022effect}. Buckling instability involves spontaneous symmetry breaking with respect to the galactic equatorial plane \citep{pfenniger1991structure, raha1991dynamical}, resulting in the thickening and weakening of bar over dynamical timescales. Observation of some galaxies having Boxy/Peanut bulge shows the presence of a bar at their center, suggesting that B/P bulge is a by-product of galactic bar formation \citep{Bettoni.Galletta.1994, kuijken1995establishing, Veilleux.etal.1999}.

Alternative mechanisms proposed to explain the development of BPX-shaped bulges include vertical resonant heating \citep{combes1990box, pfenniger1991structure, quillen2014vertical} or resonant trapping within a vILR (vertical inner Lindblad resonance) during the secular evolution of bars \citep{quillen2002growth, sellwood2020three}. Recently, \cite{10.1093/mnras/stae702, zozulia2024boxy, zozulia2025phase} investigated bar formation and thickening through the analysis of action variables and resonant angles of stellar orbits. Their results demonstrate that vertical trapping plays a dominant role in shaping the boxy/peanut morphology of galactic bars. Recent N-body simulations \citep{smirnov2018determines, sellwood2020three} have illustrated that these mechanisms can engender BP/X-shaped bulges without immediate bar buckling following bar formation.

We observe the formation of boxy/peanut (BP/X) bulges in a wide range of numerical simulations \citep[e.g.,][]{Friedli.Pfenniger.1990, raha1991dynamical, merritt1994bending, Debattista.etal.2006, Martinez-Valpuesta.etal.2006, saha2013meridional, fragkoudi2017bars, Debattista.2018, smirnov2018determines, di2019milky, Khoperskov.etal.2019, lokas2019anatomy, Smirnov.etal.2019, Collier.2020, Debattista.etal.2020, kumar2022effect}. Prior to bar formation, galaxies typically exhibit vertical symmetry. However, three-dimensional models of bar-unstable disks reveal that newly formed bars can bend out of the disk plane, losing coherence and inducing random vertical motions \citep{raha1991dynamical}. This leads to a bar that is significantly thicker than the surrounding disk and appears boxy or peanut-shaped when viewed edge-on \citep{combes1981formation, combes1990box}. Such bulges are now widely interpreted as edge-on bars. The vertical asymmetry arises after the bar formation and grows rapidly indicating a clear peak followed by rapid decreases within a few gigayears \citep{athanassoula2008boxy, lokas2019anatomy}. Recently, \cite{erwin2016caught, Li.etal.2017, Xiang.etal.2021} reported evidence of buckling instability in total eight observed galaxies, providing support for buckling driven BP/X bulges.

There have been a few attempts to study the set of galaxies having BP/X bulges \citep{ciambur2016quantifying, laurikainen2016observed, savchenko2017measuring}. However, it is not easy to detect the presence of a BP/X bulge, as it requires the bulge to be oriented in a perpendicular direction with respect to our line of sight. There has been an attempt to create a catalog of edge-on galaxies \citep{kautsch2006catalog}. Additionally, there have been attempts to study the relationship between the development of BP/X bulges and the dark matter halo using numerical simulations \citep{kumar2022effect}. Several previous studies (discussed in Section~\ref{result}) have shown a strong correlation between stellar mass and the presence of BP/X bulges. These studies have been conducted with relatively smaller sample sizes than ours. They concluded that high-mass galaxies are more likely to show the presence of a BP/X bulge than lower-mass galaxies. The gas mass fraction could play a role in determining the way in which the bar develops—either via buckling or symmetric growth—as suggested by \citet{Berentzen.etal.2007}. However, the current gas mass fraction does not appear to influence whether a BP/X bulge will form \citep{erwin2017frequency}.

Here, we aim to separate the edge-on galaxies having BP/X bulge from a large sample of edge-on galaxies obtained from the Sloan Digital Sky Survey. We classify galaxies having BP/X bulges and study different parameters like masses and sizes of galaxies with or without BP/X bulges. Then we compare their properties to find any specialty in the sample with BP/X bulges. We verified some of the past similar analysis performed previously on relatively smaller sample size. Further, we investigate the effect of BP/X hosting galaxies on the scaling laws, mass-size relation.

\section{Sample selection}
\label{sec:sample_selection}
\subsection{The data base}
\label{sec:data_base}
As a first step of our study, we aim to compile a set of edge-on galaxies as BP/X bulges are only identifiable in edge-on projection when bar is nearly side-on to our line-of-sight. For this purpose, we obtain the \texttt{ZooSpec} table from the Galaxy Zoo 1 project \citep{lintott2008galaxy, 10.1111/j.1365-2966.2010.17432.x} available with SDSS data. The Galaxy Zoo is a citizen science program where volunteer citizens help astronomers identify several unique morphological features, such as bars, spiral arms, bulge, rings etc. by voting on an interactive web interface. The \texttt{ZooSpec} table provide us with the voting data which we use to select the set of edge-on galaxies for our analysis. Followings are the two criteria we have imposed to select desired galaxies:

\begin{enumerate}
    \item[(i) ] Number of votes (\texttt{nvote}) $>10$, 
    \item[(ii)] Fraction of votes for edge-on galaxy (\texttt{p\_edge}) > 0.9
\end{enumerate}

The main motivation for choosing minimum 10 votes for each galaxy is to avoid an uncertain classification due to small number of voting. A fraction greater than 0.9 signifies that the most volunteers have identified the galaxy as edge-on. Then we obtain coordinates (RA, Dec) of the selected galaxies from the "\texttt{Galaxy}" table of the Sloan Digital Sky Survey data release 8 (SDSS DR8)\footnote{\href{http://skyserver.sdss.org/dr8/en/}{SDSS DR8}}. The "\texttt{Galaxy}" table contains photometric data of all the objects classified as galaxies from the "\texttt{PhotoPrimary}" table. Though SDSS DR17 is currently available, we use DR8 because \texttt{ZooSpec} table builds on the DR7 and its link is available with DR8 \citep{aihara2011eighth}. These conditions return us a set of 9068 edge-on galaxies. The SQL search query for this task is attached below.

\begin{verbatim}
    SELECT
    g.ra, g.dec, g.sky_r, g.petroRad_r,
    g.sky_r, g.expPhi_r
    FROM
    Galaxy g
    JOIN zooSpec z ON z.objid = g.objID
    WHERE
    z.nvote > 10 AND
    z.p_edge > 0.90
\end{verbatim}

\subsection{FITS image files}
\label{sec:fits_file}
Next, we use the (RA, Dec) of the egde-on galaxies to download the r-band images in FITS (Flexible Image Transport System) format. Our choice of $r$-band images reflects our interest in the stellar light profile of galaxies because $r$-band is less affected by dust extinction and has higher S/N ratio \citep{York.etal.2000}, which makes it easier to distinguish other structures from the oval \citep{yoshino2015box}. For this purpose, we utilize the SDSS module \footnote{\href{https://dokk.org/documentation/astropy-astroquery/v0.4.6/sdss/sdss/}{astroquery.SDSS}} of the \texttt{Astroquery} \citep{astroquery.Ginsburg.etal.2019} Python package. \texttt{Astroquery} provide a set of tools for accessing astronomical data from the webs of various surveys. To obtain the complete image of the galaxy, we use the radius of the field to be 1.5 times the Petrosian radius of the galaxy. The pre-computed Petrosian radii of the galaxies are obtained using SDSS SQL search query along with the (RA, Dec) positions as discussed in Section~\ref{sec:data_base}. For all the galaxies, we used the default SDSS pixel scale, i.e. $0.396$ arcsec/pixel. Out of 9068 galaxies, we can obtain images for 6684 galaxies because SDSS does not return images for some specific positions. Through a quick visual inspection we removed incomplete images (where only some part of the galaxy is visible) leaving a sample of 6653 galaxies for further analysis. For all the galaxies in our sample we download the corresponding PSF file from SDSS Science Archive Server.

 \subsection{Other relevant parameters}
\label{sec:other_params}
Along with obtaining FITS images of the edge-on galaxies, we also complied some relevant parameters which we will use later in this study. These parameters include object ID (objid), redshift (z), absolute magnitudes. For this purpose we make use of (RA, DEC) positions of selected galaxies in SDSS crossID tool \footnote{\href{http://skyserver.sdss.org/dr8/en/tools/crossid/crossid.asp}{SDSS crossID}}. This tool allows users to upload their own table which can be joined with different SDSS tables to extract other information using a similar SQL query as described in Section~\ref{sec:data_base}. We extract the object ID (objid), absolute magnitude in $g$-, $r$-, \& $u$-band from \texttt{Photoz} table, and redshift ($z$) from the \texttt{SpecObjAll} table. We also obtain exponential fitting parameters like minor to major axis ratio ($b/a$), effective radius (R$_{e}$), position angle, and magnitude from the \texttt{PhotoObjAll} table.

\section{FITS images processing}
\label{sec:image_processing}
As outlined earlier in the introduction, this work focuses on the characteristics of BP/X bulges and their host galaxies, we tried some well-known methods (e.g., median filtering, ellipse fitting, profile fitting, etc.) to identify the BP/X structures in our sample. In this section, we discuss the methods we tried, their drawbacks, and finally adopted method for further analysis.

\subsection{Median/Gaussian filter}
\label{sec:median_filter}
Using the median or Gaussian filter to smooth the image, and then subtracting these smoothed image from original image (and sometimes adding a factor of residual image back into the original image) to sharp the hidden BP/X structure has been used in the literature \citep{faundez1998looking}. Therefore, we first started with using the median filter available in the \texttt{Scipy} Python package\footnote{\href{https://scipy.org/}{Scipy}} \citep{2020SciPy-NMeth} to identify the BP/X hosting galaxies. The median filter smooth any image according to input box (or window) size. We give galaxy image and the box size as input parameters. The value of each pixel in the output image is calculated as the median value of the pixels inside the box of the given size around that pixel. Then we subtract the median filtered image from the original image to get the residual image. The residual image shows the salient features hidden near the central part of the galaxy as the outer regions mostly remain the same in both images and get cancel out in the residual image.

We tested this method on some images and found that the clarity of the structure in the residual image varies with the box size used for smoothing. For each image, we need to find a suitable box size by trial and error method. We could not find any specific box size that can work for all images. This method provides a work around when we want to search BP/X structures in a few images. Otherwise, it is way difficult to manually identify a correct box size for individual image given large sample like we have in this study. Therefore, we do not use median/Gaussian filter in our analysis.

\subsection{Ellipse fitting}
\label{sec:ellipse_fitting}
The subtraction of smooth intensity distribution of galaxy from its observed intensity distribution can also sharpen the hidden BP/X structure. Therefore, we construct smooth intensity distribution of galaxy by fitting iso-intensity ellipses using a Python package \texttt{Photutils} \footnote{\href{https://photutils.readthedocs.io/en/stable/api/photutils.isophote.Ellipse.html}{ellipse.photutils}} (\cite{larry_bradley_2023_7859265}). It requires an initial guess for an ellipse geometry to start the fitting. The initial ellipse is described by the user guess for center, semi-major axis, ellipticity, and position angle. We guess initial parameters to perform ellipse fitting, and use the fitted ellipses to model the smooth intensity profile of the galaxy. Then we subtract this model image from observed image to find the residual image. This residual images show the hidden structures behind the smooth intensity distribution of galaxy.

To automate this process for our large sample, we write a Python script to calculate the center of mass and perform Fourier decomposition of each image. The center of mass goes as center of ellipse, phase angle of $m=2$ Fourier mode goes as position angle, and the radius where $m=2$ Fourier mode reaches maximum goes as semi-major axis. For ellipticity ($\epsilon$), we use minor to major axis ratio ($b/a$) obtained from \texttt{Galaxy} table of SDSS data, i.e., $\epsilon=1-b/a$. Note that we would have used initial guesses (e.g., position angle and radius) from exponential fitting obtained from SDSS data for ellipse fitting \citep{stoughton2002sloan}. However, we performed Fourier analysis to minimize the failure of automatic fitting process because SDSS parameters are very different for some galaxies, e.g., some SDSS edge-on galaxies do not look like edge-on. This method works quite well as compare to median/Gaussian filter method for large sample. However, it fails fitting frequently when there is some discontinuity in the image or guessed parameters are somewhat different. For example, presence of other galaxy, bright foreground star, dust lane, etc.

\subsection{\textsc{galfit} profile fitting}
\label{sec:galfit}
Another widely used method to enhance the visibility of hidden structures is to subtract a smooth model intensity distribution from observed intensity distribution \citep{yoshino2015box}, \citep{Ness.etal.2016}. Since most of the disk galaxies follow exponential intensity distribution, we can model exponential intensity distribution using exponential fitting parameters obtained from SDSS data. However, as mentioned earlier in Section~\ref{sec:ellipse_fitting}, not all the edge-on SDSS galaxies show edge-on intensity distribution. Also, a radial exponential profile used in SDSS pipeline \citep{stoughton2002sloan} does not completely describe disk profile \citep{Kruit.Freeman.2011}. Therefore, we do not directly use the SDSS exponential fitting parameters. Instead, we performed 2D fitting using edge-on disk profile available in \textsc{galfit} 2D image decomposition tool \citep{Peng.etal.2010}. The edge-on disk profile is expressed in terms of scale radius (R$_{s}$) and scale height ($Z_{0}$) as follows \citep{van1981surface}, 
\begin{equation}
    \Sigma(R,z) = \Sigma_0 \left(\frac{R}{R_s}\right) K_1\left(\frac{R}{R_s}\right) \text{sech}^2\left(\frac{Z}{Z_0}\right),
    \label{eqn:edge_disk}
\end{equation}
where $\Sigma_0$ is the central surface brightness, $K_1$ is a Bessel function.

\textsc{galfit} is a stand-alone 2-D image decomposition program written in C language. It uses an arbitrary number and mix of parametric functions (such as Sersic, Nuker, Gaussian, Moffat, and exponential) to decompose the image. It creates the model image and iterates it over and over the original image to find the perfect fit. It is capable of fitting galaxies and the sub-structures in them. Although the presence of dust significantly affects photometric decomposition in edge-on galaxies, its impact must be carefully considered. Using radiative transfer modeling of simulated galaxies, \cite{2023MNRAS.524.4729S} showed that dust can lead to a 25–50\% increase in the fitted disk scale radius in edge-on systems. In contrast, we expect bar formation to enhance central brightness, which typically results in a reduced disk scale radius in single component fitting, while bar thickening is expected to produce counter effect similar to presence of the dust. A detailed investigation of the counterbalancing effects of dust and bar presence is beyond the scope of this paper. In the case of extensive data like in this study, automation is very essential. For automation, user need to provide an external `wrapping' algorithm for taking care of the pre-processing and post-processing of the fittings. For our image processing, we provided an input file containing all the input parameters (initial guesses) to start the computation. Next, we discuss these initial guesses.

\subsection*{Initial guesses}
To initialize the fitting process in \textsc{galfit} we need to provide initial guesses for central surface brightness, scale radius and scale height in equation~(\ref{eqn:edge_disk}) along with center and position angle. We obtain the initial guesses for center, position angle and scale radius using center of mass and Fourier analysis as discussed in Section~\ref{sec:ellipse_fitting}. Being the sample of edge-on disk galaxies, the initial guess for scale height is assumed to be 0.2 $\times$ scale radius. If fitting fails, we try with larger scale height of 0.5 $\times$ scale radius, particularly when galaxies do not look like edge-on. Since \textsc{galfit} performs edge-on disk surface brightness fit in $mag/arcsec^{2}$. To guess the central surface brightness, we consider a square of 10 pixels $\times$ 10 pixels size around the center of galaxy, and measure total counts in this box as ADUs. Then, we measure the central surface brightness in $mag/arcsec^{2}$ using following expression,
\begin{equation}
    \Sigma_{0, \text{guess}} = -2.5 \log \left(\frac{ADUs}{\Delta x\Delta y ~t_{exp}}\right) + mag_{zpt},
\end{equation}
where $\Delta x$ and $\Delta y$ are the side lengths of the region in x and y direction respectively, here taken as 10 pixels in each directions. Time $t_{exp}$ is the exposure time, which is taken to be 53.9 seconds for the r-band. Magnitude $mag_{zpt}$ is the zero point magnitude, here taken to be 24.8 \citep[see Table 21 of][]{stoughton2002sloan}.

\begin{figure*}
   \centering
    \includegraphics[width=\linewidth]{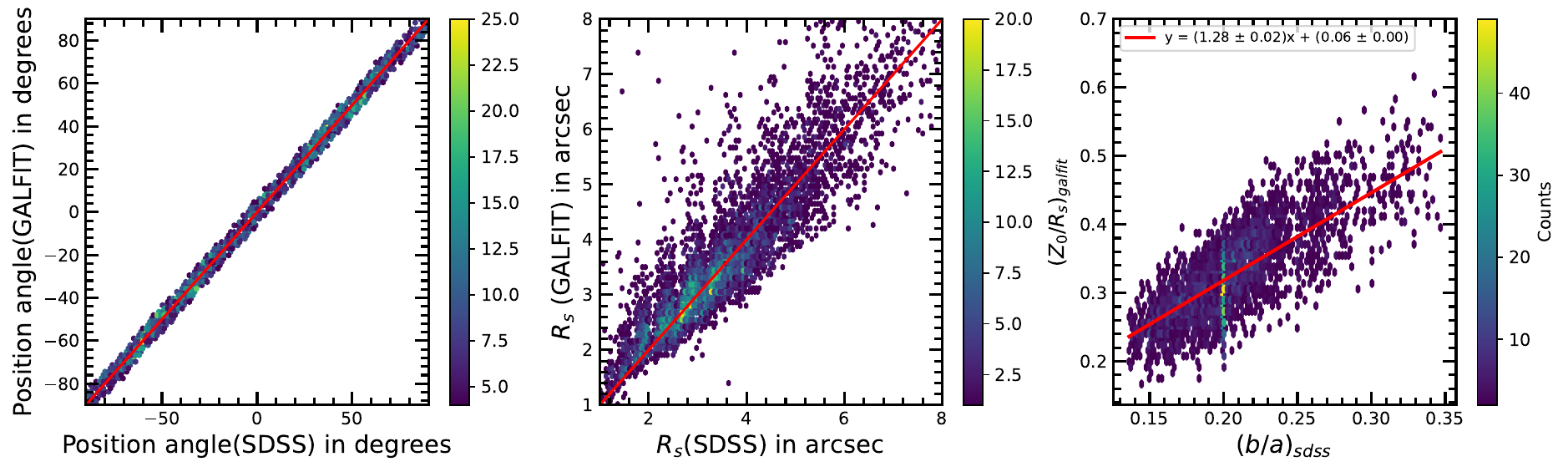}
    \caption{Figure showing the comparison of fitted position angle, radius and axial ratio between edge-on disk profile fitting by \textsc{galfit} and exponential disk profile fit parameters in SDSS. The red line in left most and middle panel gives the line corresponding to $y=x$. In the right most panel the fitting line is given in red with the fitting parameters. $Z_0$ is the disk height and $R_s$ denotes the scale radius obtained from \textsc{galfit}.}
    \label{fig:galfit_sdss_param} 
\end{figure*}

For each galaxy, all these initial guess parameters are compiled into a text file in \textsc{galfit} input format. It also includes galaxy name, output file name, convolution box size among other parameters. Our wrapping algorithm for automation takes care of the variable parameters, e.g., initial guesses, galaxy name, output file name, etc. This fitting analysis provides us with edge-disk parameters which are not available in the original SDSS catalog. In Fig.~\ref{fig:galfit_sdss_param}, we show comparison of exponential disk parameters obtained from SDSS and edge-disk parameters from our \textsc{galfit} decomposition. The left panel compares position angles, and the middle panel compares scale radii. Our analysis shows a good agreement between SDSS and \textsc{galfit} position angles and scale radii as evident from the equality line displayed in left most and middle panels.
The SDSS exponential fitting semi-minor to semi-major axis ratio ($b/a$) and \textsc{galfit} edge-disk fitting scale height to scale radius ratio ($Z_{0}/R_{s}$) exhibits linear relation but non-unity slope as shown in the right most panel. We fit the straight line and obtain following relation,
\begin{equation}
    \left(\frac{Z_{0}}{R_s}\right)_{\text{galfit}} = (1.28 \pm 0.02) \left( \frac{b}{a}\right)_{\text{sdss}} + 0.06.
\end{equation}

\section{BP/X bulge classification}
\label{sec:classification}
After edge-disk fitting using \textsc{galfit}, we visually inspected residual images ($=$ original image $-$ model image) of all 6653 galaxies to build a catalog of galaxies with and without BP/X bulges for detailed study. The sample without BP/X bulges is considered as control sample for comparison with the sample having BP/X bulges. We identify BP/X bulges in different stages/contrast. Following is our classification scheme for BP/X bulges and control sample.

\begin{figure*}
    \centering
    \includegraphics[width=\textwidth]{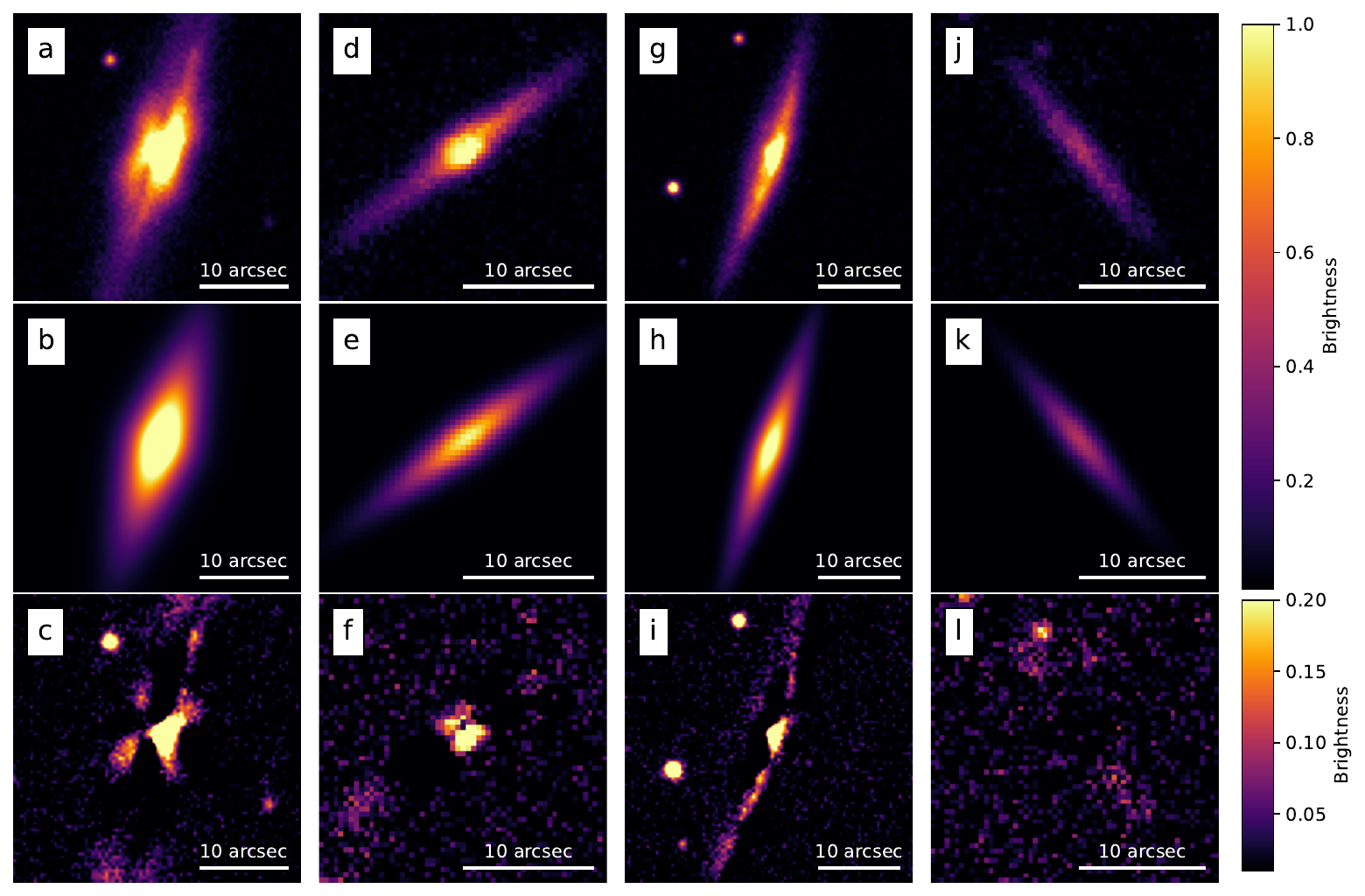}
    \caption{The figures in the first row are four galaxies in our sample. The second row shows the model fitted using \textsc{galfit} for the corresponding galaxy above it. The third row gives the residual image of the corresponding galaxy above it. Each column of figures gives an example of galaxies classified as strong both-sided, weak both-sided, one-sided, and control, respectively from left to right. The white scale bar marks the scale of 10 arcsec.} 
    \label{fig:bpx_classification}
\end{figure*}

\begin{itemize}
    \item Both-sided:- This class of galaxies show the presence of boxy, peanut, or x-shape structure at their center in the residual images.
    The structure can be clearly visible as in Fig.~\ref{fig:bpx_classification}(a,b,c) or can be faintly visible as in Fig.~\ref{fig:bpx_classification}(d,e,f). We identify 1673 galaxies in this class, which comprise nearly $25\%$ of the total sample. On the basis of visibility, this class can be further divided into strong and weak both-sided structures. In our analysis, we refer to both-sided as the collection of both strong and weak both-sided structures, unless specified differently.
    
    \item Strong both-sided:- From the class of both-sided BP/X bulge galaxies, we separate galaxies showing a clear x-shape structures and classify them into strong both-sided sample. These galaxies show a very prominent x-shape structure as in Fig.~\ref{fig:bpx_classification}(a,b,c). There are 504 galaxies (nearly $8\%$) in this class.

    \item One-sided:- The residual images of these galaxies shows excess intensity on one side of the galactic mid-plane. It seems like the bulge is present on only one side of the galaxy. Fig.~\ref{fig:bpx_classification}(g,h,i) is an example of the galaxy in this sample. One of the possible reason for this appearance is the dust obscuration on one side. Additionally, it is also possible that this class potentially contains galaxies in the initial buckling phase. There are 1112 galaxies in this class, which is nearly $17\%$ of the total sample.

    \item Control:- These galaxies show no sign of BP/X bulge formation. Therefore, the model completely fits the original image, leaving no feature in the residual. There are a total of 3868 galaxies in this class. Some of the galaxies in the control sample show presence of other significantly bright structure in the projection, e.g., in the bottom left side of Fig.~\ref{fig:bpx_classification}(j,k,l). There are 146 galaxies which show nearby bright structures (likely neighbor galaxy). The remaining 3722 galaxies do not show any neighboring bright objects in the projection. So, nearly $58\%$ of our sample belongs to this class. Note that some of the galaxies showing showing spheroidal structures (classical bulge or end-on projected bars) are also included in the control sample.
\end{itemize}      

For more examples of different categories of BP/X bulges discussed here, we refer the reader to Fig.~\ref{fig:more_strong_bpx}, \ref{fig:more_weak_bpx}, and \ref{fig:more_one_sided} in Appendix~\ref{app:more_bpx_example}. We note that the weak both-sided and one-sided structures may contain some oblate classical bulge structures, which are difficult to distinguish due to spatial resolution limit of the SDSS. We caution the user to consider this caveat when interpreting results for these two classes of BP/X structures in this study.

\section{Derived quantities for galaxies}
\label{sec:derived_quant}
Along with the parameters obtained from SDSS catalog and measured using 2D image decomposition using \textsc{galfit}, we derive stellar masses, distances and physical sizes of the galaxies from existing information. The following subsections describe our methodology for measuring these derived properties of galaxies.

\subsection{Stellar masses}
\label{sec:stellar_mass}
To proceed with the stellar mass estimation, we require a mass-to-light ratio relation. For this purpose, we use a widely adopted stellar mass-to-light ratio relation available for SDSS filters.
\begin{equation}
    \log{\gamma^{j}} = a_{j} + (b_{j} \times color),
\end{equation}
where $\gamma^{j}$ is the mass-to-light ratio of the galaxy in $j^{th}$ imaging band. Parameters $a_{j}$ and $b_{j}$ are two constants for the same band at a given color index (here, $g-r$). We have obtained these constants from \cite{bell2003optical, du2020self}. Both, \cite{bell2003optical} and \cite{du2020self} report similar $a$ and $b$ parameters for the $g-r$ color. Therefore, we consider using $g-$ and $r-band$ magnitudes for the stellar mass estimation of our sample galaxies \citep{kumar2022growth}. We obtain the galactic extinction corrected $r$- and $g$- band magnitude from SDSS \texttt{PhotoObjAll} table. The internal extinction in $r$- band is calculated using the relation $A_r = \gamma_r~\log_{10}(a/b)$, where $\gamma_r = 1.37$ is obtained from \cite{shao2007inclination}. Using the relation $\frac{A_g}{A_r} \approx \frac{R_g}{R_r} \approx 1.4$ \citep[see Table 2 of][]{yuan2013empirical}, we measure the internal extinction in $g$-band. Then the internal extinction corrected color and magnitude are given as $$(g-r)_{corr} = (g-r)_{obs}-(A_g-A_r), \quad M_{corr} = M_{obs}-A_r.$$ We use the extinction corrected absolute magnitudes and colors for mass calculation. 

After calculating mass-to-light ratio using above described method, we use absolute $r$-band magnitudes (M$_{r}$) to measure the luminosities of galaxies in solar unit by following expression.
\begin{equation}
    L = 10^{-\frac{M_{r} - M_{\odot}}{2.5}},
\end{equation}
where, M$_{\odot}$ is solar magnitude in $r$-band. For our analysis, we obtained solar magnitude from \cite{blanton2003galaxy}. Next, we measure stellar mass of the galaxies in solar mass unit by multiplying mass-to-light ratio with their luminosities as follows,
\begin{equation}
    \text{Stellar mass} = \gamma^{j} \times L.
\end{equation}

\subsection{Angular distances and physical sizes}
\label{sec:distance}
We use the spectroscopic redshifts of galaxies obtained from SDSS database to measure their angular distances. In the flat universe, the angular distance of an object at redshift $z$ is given by the following expression \citep{simard2011catalog}.
\begin{equation}
    D_a = \frac{c}{H_0(1+z)} \int_{0}^{z} \frac{1}{\sqrt{\Omega_m(1+x)^3 + \Omega_\Lambda}} \,dx,
\end{equation}
where $c$ is the speed of light in vacuum, $H_0$ is the Hubble's constant, $\Omega_m$ is the matter density of the universe, and $\Omega_\Lambda$ is the dark energy density. For this study, we have considered $H_0 = 70 \,km \,s^{-1} \,Mpc^{-1}$, $\Omega_m = 0.3$, and $\Omega_\Lambda = 0.7$. The peculiar velocities of nearby galaxies affect the redshift measurement, and thus their distance using above expression. Nonetheless, all the galaxy in our sample have $z > 0.005$, indicating marginal effect of peculiar velocities on the distance estimation \citep{Shen.etal.2003}.

From the \textsc{galfit} photometric decomposition, we obtain the disk scale height and disk scale radius in pixels. Using the default SDSS pixel size of 0.396 arcsec/pixel,  we calculate the angular size of galaxies in radian. Then we obtained the physical sizes of galaxies by multiplying angular sizes and angular distances.

\section{Sample characteristics}
\label{sec:sample_characteristics}
\begin{figure*}
   \centering
    \includegraphics[width=\linewidth]{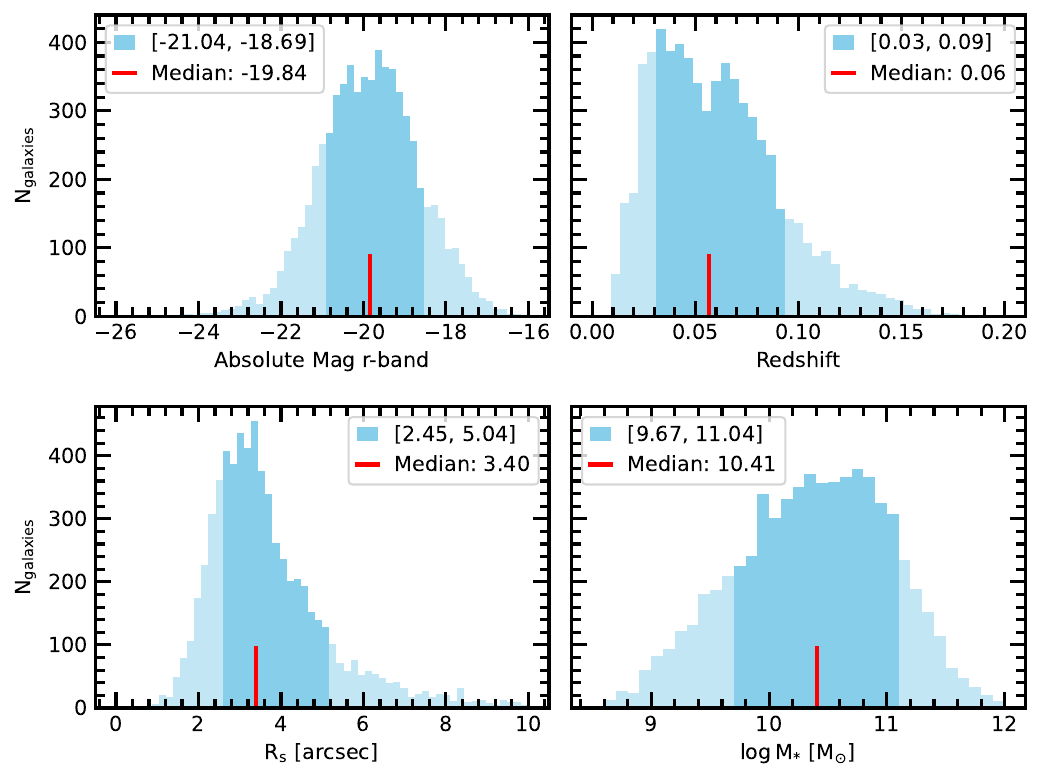}
    \caption{Absolute $r-$band magnitude, redshift, angular scale radius, and mass distribution of galaxies in our base sample. The red color vertical lines correspond to median values, and the dark shaded regions show 16$th$ to 84$th$ percentiles range of the distributions.}
    \label{fig:sample_characteristics} 
\end{figure*}

As mentioned in Section~\ref{sec:fits_file}, we started with 6684 edge-on galaxies, and performed 2D morphological decomposition using \textsc{galfit}. Further, we visually separated these objects into four different subclasses depending on central features as discussed in Section~\ref{sec:classification}. Fig~\ref{fig:sample_characteristics} shows the distributions of magnitudes, redshifts, scale radii, and stellar masses of our complete sample. The median values of the distributions are represented with the red color vertical lines at the bottom axes of respective panels. The lower and upper limits of dark distributions respectively indicates $16\%$ and $84\%$ range as shown in the legends of respective panels.

Top left panel of Fig.~\ref{fig:sample_characteristics} shows the absolute magnitude distribution of the sample galaxies. Our sample cover an absolute $r-$band magnitude from $-16$ to $-26$. The median of magnitude distribution lies at $-19.8$ with $1\sigma$ percentile range from -18.6 to -21.0. As most of the galaxies are bright and resolved, they are primarily local volume objects. The median redshift of the sample is 0.06 with $1\sigma$ percentile range from 0.03 to 0.09 as can be seen from top right panel of the figure. The farthest galaxy in the sample lies at redshift 0.2.

The bottom left panel of Fig.~\ref{fig:sample_characteristics} displays the disk scale radius of galaxies. The median angular scale radius of the sample is $3.4~arcsec$ with $16-84\%$ range $2.4-5.0~arcsec$ in $r-$band. A few nearby galaxies show larger angular size greater than $10~arcsec$. For clarity, we have shown the distribution up to $10~arcsec$. Our derived stellar masses of the galaxies span a wide range from $2 ~\times ~10^{8}$ to $5 ~\times ~10^{12} ~M_{\odot}$. The median stellar mass of the sample is $2.5 ~\times ~10^{10} ~M_{\odot}$ and the $1\sigma$ percentile range covers $4.6 ~\times ~10^{9}$ to $1.2 ~\times ~10^{11} ~M_{\odot}$ (see the bottom right panel of Fig.~\ref{fig:sample_characteristics}).

Since the identification of BP/X structures is affected by the inclination of galaxies, it is worth discussing the inclination distribution of our sample. We measure the inclination angle using the axis ratio $b/a$ obtained from SDSS data, applying the relation $ \cos(i) = \sqrt{((b/a)^2 - q^2) / (1 - q^2)} $ \citep{1958MeLuS.136....1H}, where $q$ is the intrinsic axial ratio. Following \cite{1977A&A....54..661T}, we adopt $q = 0.2$ for this analysis. We find a median inclination of $\approx 87$ degree, with a $16$–$84\%$ range of $79$–$90$ degrees. For an ideally edge-on galaxy, the inclination angle is $90$ degree, indicating that our sample largely consists of nearly edge-on galaxies.

\section{Result}\label{result}
\subsection{Mass and size distributions of BP/X bulges}
\label{sec:bulge_mass_size}
\begin{figure*}
   \centering
    \includegraphics[width=\linewidth]{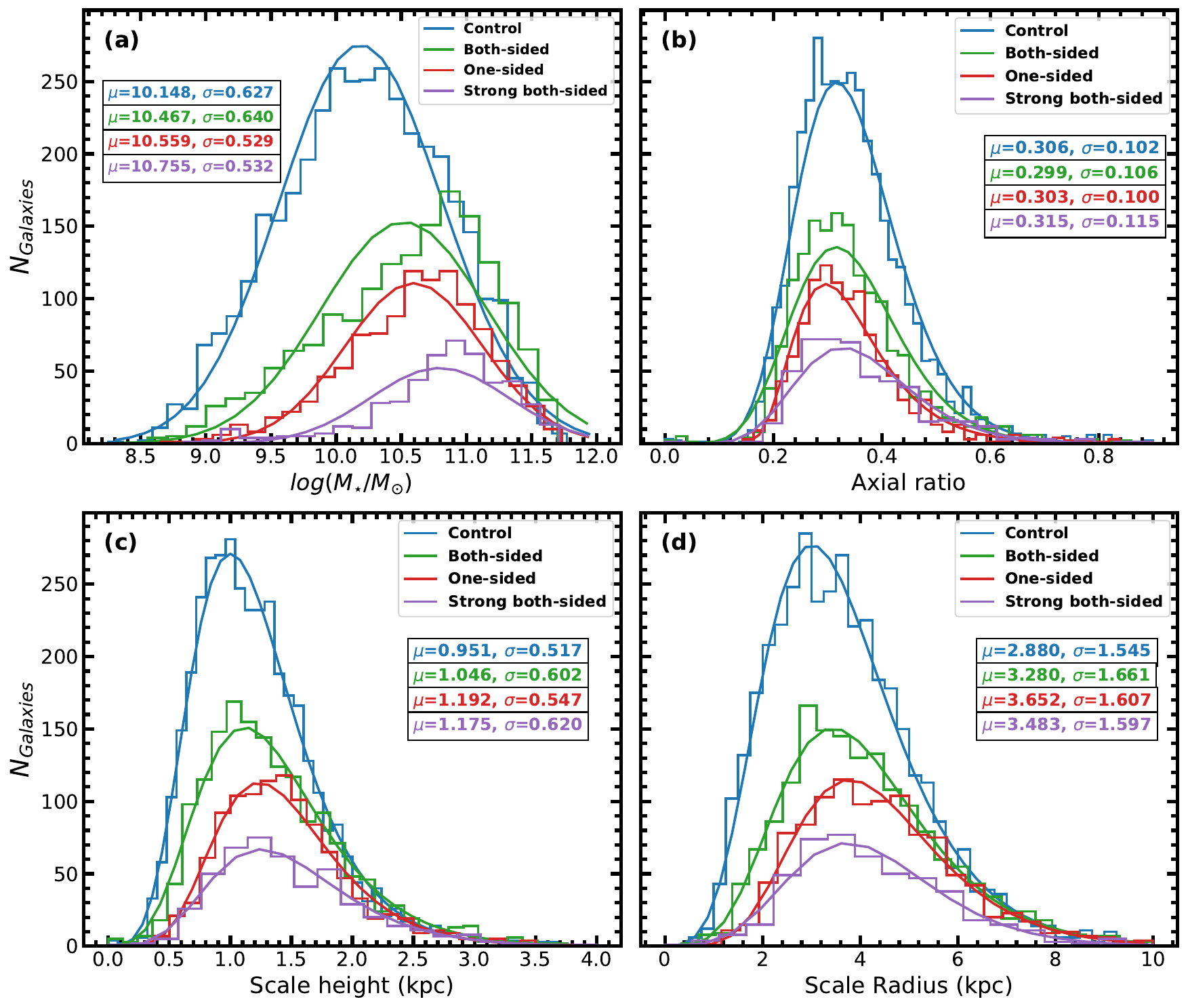}
    \caption{Distribution of different parameters along with the `log norm' fitting parameters for the distributions of mass, axial ratio, scale height and radius for different samples classified based on the presence of BP/X bulge in them.}
    \label{fig:bulge_mass_size} 
\end{figure*}

First, we investigate the differences in properties like mass, disk radius, disk height and axial ratio among the different classes we have identified in the whole sample. In Fig.~\ref{fig:bulge_mass_size}, we plot the distributions of mass, axis ratio, disk height, and disk radius for the four samples in this study. The control, both-sided, one-sided, and strong both-sided are respectively shown with blue, green, red, and purple colors. To better understand the distribution, we fitted each dataset using the `log normal' distribution. For fitting, we used the Python package \texttt{Scipy}.

The distribution of masses (\ref{fig:bulge_mass_size}a) shows that the mass distribution for the sample without any signature of bulge formation (control sample) peaks at $\log (M_{*}/M_{\odot}) \approx 10.1$. But for the other 3 samples, i.e. the ones showing signatures of bulge formation peaks in the range of $\log (M_{*}/M_{\odot}) > 10.5$. This is expected because bulge formation occurs in galaxies having a bar and high mass galaxies are more likely to host bars than the lower mass galaxies (\cite{2018MNRAS.474.5372E}).

We have defined axial ratio as the ratio of disk scale height to disk scale radius. For the galaxies having signature of bulge formation, we expect the axial ratio to be higher than that of the control sample as bars thicken during BP/X bulge formation. From our `lognorm fit', we obtained the mean value of axial ratio of strong both-sided galaxies higher among all samples (\ref{fig:bulge_mass_size}b), though not significantly different. This similarity in the axial ratio is also evident from the scale height (\ref{fig:bulge_mass_size}c) and scale radius distributions (\ref{fig:bulge_mass_size}d).

For further analysis described in Section \ref{sec:stellar_surface_density} and \ref{sec:effect_of_bpx_on_galaxy_evolution}, we have used only the galaxies having stellar mass, disk radius and disk height less than $5 \times 10^{11} M_\odot$, 14~kpc, and 5~kpc, respectively. By using these limits, we remove relatively higher value outliers in the data. These limits do not affect the correctness of the analysis because the sample includes nearly 98.3\% of galaxies. Category-wise, we have nearly 99\% of control sample and one-sided sample, nearly 97.4\% of both-sided and 95.4\% of strong both-sided within this limit.

\subsection{BP/X fraction}
\label{sec:bpx_fraction}
There have been a number of attempts to estimate the BP/X bulge fraction for different-size samples of edge-on galaxies. The earliest study was by \cite{jarvis1986search}, who estimated BP/X frequency of 1.2\% in his relatively small sample of 41 disc galaxies. In a sample of 117 visually edge-on galaxies with diameter $D_{25} \ge 3.5^{'}$, \cite{shaw1987nature} estimated that approximately 20\% contain a BP/X bulge. He noted this as a conservative lower limit, as additional BP/X bulge galaxies were identified that did not appear in the primary masking. Using a larger sample of 555 galaxies with axial ratio $<0.5$, \cite{de1987box} estimated the BP/X bulge fraction to be 13\%. 

\cite{lutticke2000box} presented the first detailed statistics of BP/X bulges using a large sample of $\sim 1350$ edge-on galaxies from Third Reference Catalog of Bright Galaxies \citep[RC3:][]{de_Vaucouleurs.etal.1991}. They found $45\%$ BP/X bulges in their sample of 734 classifiable bulges. \cite{yoshino2015box} decomposed about 1700 SDSS DR7 disc galaxies, and identified BP/X structure. In their clean sample of 1329 galaxies in  $i-$band, they reported $22\%$ BP/X structures. Since \cite{yoshino2015box} used SDSS sample, we cross-matched our sample with the their catalog of 1312 edge-on galaxies in the $r$-band, and identified 455 galaxies in common. The primary reason for the discrepancy in sample sizes lies in the differing selection criteria.\cite{yoshino2015box} selected galaxies with a minor-to-major isophotal axis ratio in the $r$-band less than 0.25, whereas our sample was selected based on volunteer votes. When we compare the $b/a$ values from their model fitting for both the common and complete samples, we find that the common sample predominantly consists of galaxies with $b/a < 0.2$, while complete sample show median $b/a \approx 0.2$, indicating that a large fraction of their sample comprises moderately inclined systems. Nonetheless, out of these common galaxies they identified 129 total BP/X structures (7 strong, 42 standard, and 80 weak in their definitions) and we identified 165 total BP/X structures (68 strong both-sided and 97 both-sided). Recently, a similar fraction of BP/X bulges is reported in $r$-band images. \cite{Marchuk.etal.2022} trained Artificial Neural Network (ANN) to automatically identify BP/X structures in large sample of 13048 edge-on galaxies obtained from DESI Legacy Imaging Survey \citep{Dey.etal.2019}. They found about 1925 ($\sim 15\%$) galaxies with BP/X structures.

As described previously in Section~\ref{sec:classification}, we started with 6653 edge-on galaxies and classified them into three classes of BP/X bulges: Both-sided, Strong both-sided, and One-sided. We identify 1673 ($\sim 25\%$) both-sided bulges, and 1112 ($\sim 17\%$) one-sided bulges. Among both-sided bulges 504 ($\sim 8\%$) are strong both-sided with prominent X-structure, and remaining 1169 ($\sim 17\%$) show faint X-structure. When combined together, both-sided and one-sided make $\sim 42\%$ (2785 galaxies) of the total sample. Though our base sample is nearly half of that studied in \cite{Marchuk.etal.2022}, we have visually verified the presence of BP/X structure in each galaxy by manually varying intensity of the residual images to avoid missing any faint structure. Therefore, our analysis presents largest and complete sample of BP/X structures till date with detailed classification.

\subsection{Relationship of BP/X bulge with the stellar mass}
\label{sec:relationship_with_stellar_mass}
The fraction of barred galaxies shows a strong correlation with the stellar mass of a galaxy \citep{Nair.Abraham.2010, Melvin.etal.2014, Sodi.2017}. Generally, massive galaxies tend to have bar more likely. Analogously, by considering a sample of 84 barred galaxies, \cite{erwin2017frequency} showed that the BP/X bulges are more common in galaxies with higher stellar masses. They found about 80\% of barred galaxies with stellar masses $\rm > 10^{10.4} \,M_{\odot}$ have BP/X bulges. They concluded that the frequency of BP/X bulge in barred galaxies is nearly 50\% in the stellar mass range $\log(M_{*}/\Msun) \approx 10.3-10.4$. The similar stellar mass-dependence of BP/X bulges has been reported in other studies \citep{Marchuk.etal.2022}, and at higher redshift as well \citep{Kruk.etal.2019}.

In Fig.~\ref{fig:bulge_mass_size}a, we have plotted the stellar mass distribution of both-sided (green), strong both-sided (purple), one-sided (red), and control sample (blue) in this study. We found that the distribution of control galaxies peaks at lower stellar mass than the galaxies with BP/X bulges. The median of the control sample lies at  $\log(M_{*}/\Msun) = 10.14$, whereas the sample with both-sided, strong both-sided, and one-sided BP/X bulges have median at $\log(M_{*}/\Msun) = 10.46, 10.75, and\ 10.55$, respectively. Our results are consistent with the previous findings of \cite{erwin2017frequency, Kruk.etal.2019, Marchuk.etal.2022}, suggesting massive galaxies most likely host the BP/X bulges.

\begin{figure}
    \centering
    \includegraphics[width=\linewidth]{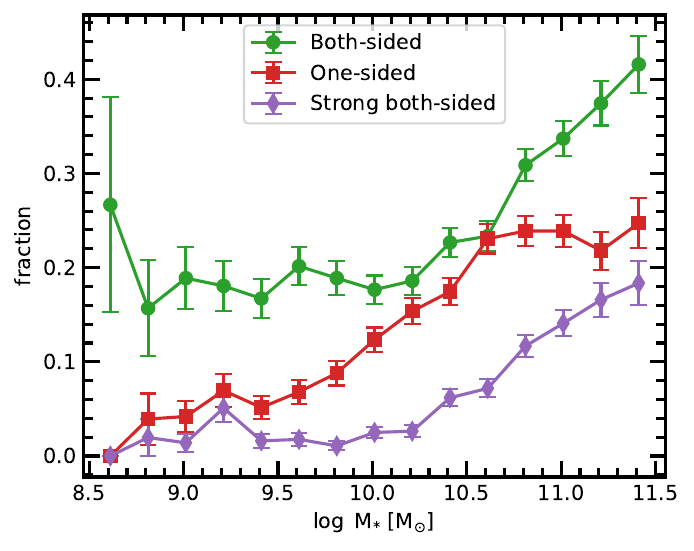}
    \caption{Fraction of galaxies hosting both-sided (green), one-sided (red) and strong both-sided (purple) BP/X structures as a function of their stellar mass. Fraction of BP/X hosting galaxies increases with their stellar mass.}
    \label{fig:bpx_frac_with_mass}
\end{figure}

Further, we measure the BP/X fraction in our sample of edge-on galaxies. Fig.~\ref{fig:bpx_frac_with_mass} shows the fraction of both-sided (green), strong both-sided (purple), one-sided (red) bulges with the stellar mass of galaxies. The vertical bars in respective curve represent statistical uncertainty in measurement defined as $\sqrt{f \times (1-f) / N}$, where $f$ being fraction of a bulge class at any stellar mass bin and $N$ being total number of galaxies in that bin. Similar to the previous studies, we notice increasing BP/X bulge fraction with increasing stellar mass of galaxies. In the stellar mass range shown here, the fraction of one-sided BP/X bulges is always higher than the strong both-sided bulges at any given stellar mass. Additionally, the fraction of both-sided BP/X bulges, which also includes strong both-sided bulges, is always the highest for any stellar mass. In the shown stellar mass range, we can see an increase in strong both-sided, one-sided and both-sided bulge fraction, respectively, from 0, 0, and 20 to 18, 26, and 42 percent.

\subsection{Dependence of BP/X bulge on stellar surface density}
\label{sec:stellar_surface_density}
\begin{figure}
    \centering
    \includegraphics[width=\linewidth]{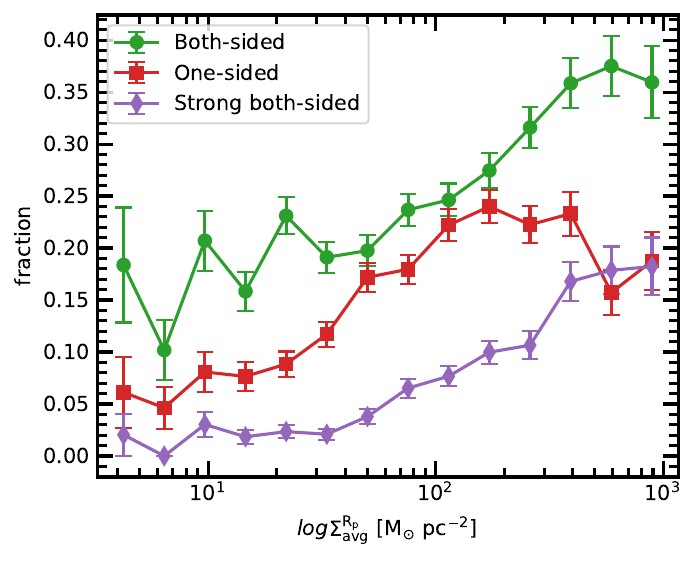}
    \caption{The fraction of both-sided (green), one-sided (red), and strong both-sided (purple) BP/X structures as a function of average surface density of galaxies. The high surface density disks host more BP/X structures than the low-surface density disks.}
    \label{fig:bpx_frac_with_dens}
\end{figure}

To investigate the dependence of BP/X structures formation on the disk stellar surface density, we measure the average stellar surface density of galaxies in our sample. We consider that the total stellar mass of a galaxy lies within its Petrosian radius and define average stellar surface density as $\rm \Sigma^{R_{p}}_{avg} = M_{*}/(\pi R_{p}^{2})$. Note that we consider Petrosian radius for measuring average stellar density instead of scale radius because measurement of scale radius is significantly affected by the secular evolution of inner disk and presence of dust as mentioned earlier in Section~\ref{sec:galfit} (see Appendix~\ref{app:eff_of_scale_radius} for similar analysis using scale radius). In Fig.~\ref{fig:bpx_frac_with_dens}, we compare the fraction of both-sided, one-sided, and strong both-sided BP/X structures as a function of average stellar surface density of galaxies, respectively, in green, red, and purple colors. 

Fig.~\ref{fig:bpx_frac_with_dens} clearly shows that the fraction of any type of BP/X structure increases with the increasing average surface density of galaxies. The low surface density disks ($\rm \Sigma^{R_{p}}_{avg} \lesssim 20~\MsunPc$) show nearly constant fraction of both-sided and strong both-sided BP/X bulges. Below this surface density limit, the fraction of strong both-sided structures is less than $3\%$, while both-sided makes about $15-20\%$, and the contribution of one-side structures is about $6-8\%$. For high surface density disks ($\rm \Sigma^{R_{p}}_{avg} \gtrsim 20~\MsunPc$), the fraction of BP/X structure hosting disks rises by about $15-20\%$ in all three groups. This analysis shows that the high surface density disks are more likely to host BP/X structures as compare to the low surface density disks.

The increasing fraction of BP/X structures with increasing average surface density of galaxy provides additional constraint on the BP/X bulge formation. Though the fraction of BP/X structures strongly depends on the stellar mass of galaxies, this study suggests the domination of BP/X structures in high surface density disks. It is consistent with the bar formation criterion that the high surface density disk have low Toomre-Q parameter, which makes disk unstable against dynamical instability. An unstable dynamically cold disk can form bar easily, which can eventually thicken to form a BP/X structure \citep{combes1981formation, raha1991dynamical, athanassoula2005nature}.

\subsection{Effect of BP/X structure on galaxy evolution}
\label{sec:effect_of_bpx_on_galaxy_evolution}

\begin{figure}
    \centering
    \includegraphics[width=\linewidth]{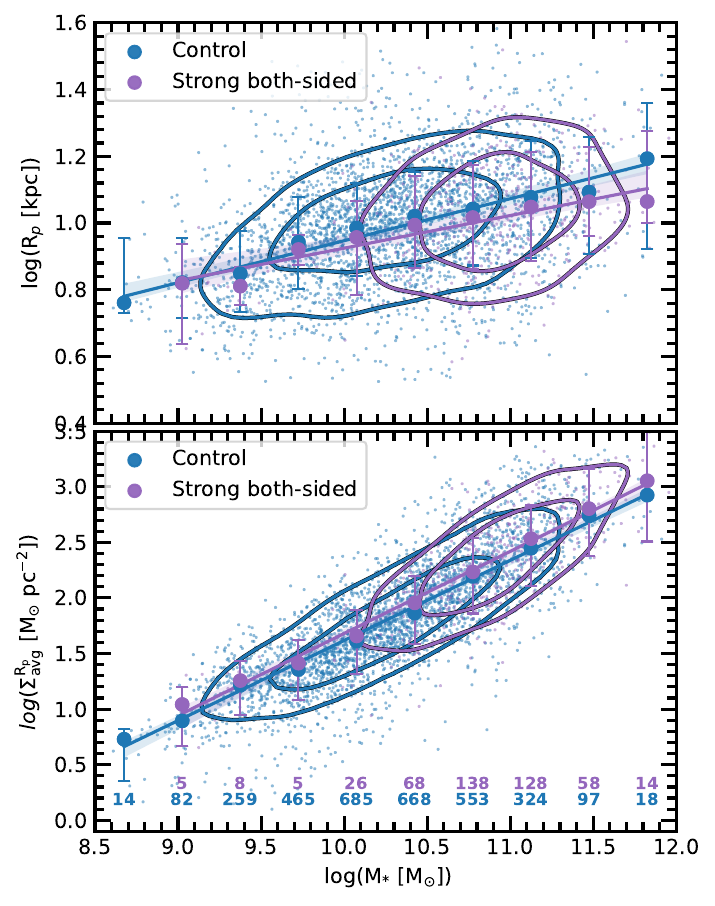}
    \caption{The stellar mass-size relation (top panel) and stellar mass-stellar surface density relation (bottom panel) of galaxies hosting BP/X structures. Control and strong both-sided samples are shown in blue and purple colors, respectively. The high-mass disks having strong both-sided BP/X structures are compact and contribute to scatter in these scaling relations at high-mass end.}
    \label{fig:log-relation}
\end{figure}

\subsubsection{The mass-radius relation}
\label{sec:mass_size_relation}
The mass-radius relation of galaxies is a fundamental scaling law that links a galaxy’s stellar mass to its physical size.  This relation shows that more massive galaxies tends to be larger in size. However, the slope and scatter of the relation depend on galaxy type. Early-type galaxies (ellipticals and lenticulars) typically exhibit a steeper mass–radius relation, indicating more compact structures at fixed mass compared to late-type (disk-dominated) galaxies \citep{Shen.etal.2003, van_der_Wel.etal.2014}. At a given stellar mass, early-type galaxies are smaller than late-type galaxies, reflecting differing formation mechanisms - such as gas-poor mergers for ellipticals and inside-out star formation for disks \citep{Naab.etal.2009, Nelson.etal.2016}.  Moreover, high-redshift galaxies ($z \gtrsim 2$) are observed to be significantly more compact at fixed stellar mass than their local counterparts, suggesting substantial size growth over cosmic time \citep{Trujillo.etal.2006, van_Dokkum.etal.2010}. Understanding the mass–radius relation is therefore critical for constraining galaxy assembly histories and testing models of galaxy evolution.

To investigate the role of galaxies hosting BP/X structures---and consequently the role of barred galaxies---in the mass–size relation of galaxies, we plot the Petrosian radius as a function of stellar mass for control sample and galaxies having strong both-sided BP/X bulges in the top panel of Fig.~\ref{fig:log-relation}. We did not include both-sided and one-sided sample in this analysis to avoid any bias due to presence of heavy dust lanes, which may lead to their wrong classification in some cases. Also, we use Petrosian radius instead of scale radius to minimize the effect of dust obscuration and bar evolution on scale radius size. The control and strong both-sided samples are shown using blue, and purple color open circles, respectively. We have also plotted density contour for both the samples to show the spread of data using their respective colors. To show the size evolution with mass we have plot the median sizes in different mass bins along with 16th and 84th percentile ranges using large filled circles. The number of galaxies in each bin are listed at the bottom of the bottom panel figure. The straight lines are fit to median data points.
To avoid the effect of redshift evolution, we have considered only those galaxies which lie within $1\sigma$ of the median redshift of our sample, i.e., $z=0.03$ to 0.09. We find that galaxies in the strong both-sided sample are typically smaller in size as compare to galaxies in control. Additionally, the disks with strong both-sided sample show shallower relation than control sample. This shallow trend contribute to large scatter at high-mass end of mass-size relation. {However, the scatter in the sample is very high compare to the difference between to samples.}

\subsubsection{The mass-density relation}
\label{sec:mass_density_relation}
In analogous to mass-radius relations, the mass-density relations shows that more massive galaxies tend to have higher surface density. The early-type galaxies are more compact and centrally concentrated than the late-type galaxies. This correlation reflects the underlying processes governing star formation efficiency, feedback, and mass assembly. For instance, compact massive galaxies at high redshift exhibit high stellar surface densities, suggesting rapid, early star formation followed by quenching and morphological transformation \citep{Barro.etal.2013}.

We study the effect of BP/X hosting galaxies on the mass-density relation of spirals by fitting a straight line in log-log scale similar to the mass-size relation and plot in the bottom panel of Fig.~\ref{fig:log-relation}. Again, blue color shows control sample and purple represents strong both-sided bulges. The mass-density relations are significantly stronger than the mass-size relations and shows higher degree of correlation. The galaxies having strong both-sided BP/X structures are typically compact towards high-mass end of the relation. The mass-density relation does not change as significantly as mass-size relation of galaxies.

\section{Summary}
In this paper, we performed a comprehensive study of BP/X structures in edge-on galaxies obtained from SDSS DR8. We compiled an initial sample of 6684 galaxies with more than 0.9 probability of being edge-on based on more than 10 votes in GalaxyZoo survey. We used GALFIT for 2D image decomposition to measure the scale radii and scale heights of galaxies. Further we used the residual images from 2D decomposition to manually identify BP/X structures in each image and categorized the sample in four classes depending on the signature of BP/X structure$--$(1) strong both-sided: galaxies with clear BP/X structures, (2) both-sided: galaxies with faint or clear BP/X structures, (3) one-sided: galaxies with excess flux on one side on the galactic plan, and (4) control: galaxies without BP/X structures. Additionally, We calculated the mass, distance, and physical size of galaxies. Followings are key summarizing points of our study.

\begin{itemize}
    \item We visually classified 6653 edge-on galaxies into 1673 ($\sim 25\%$) both-sided bulges, and 1112 ($\sim 17\%$) one-sided bulges. Among both-sided bulges 504 ($\sim 8\%$) are strong both-sided with prominent X-structure, and remaining 1169 ($\sim 17\%$) show faint X-structure. When combined together, both-sided and one-sided make $\sim 42\%$ (2785 galaxies) of the total sample. Our sample provides the largest catalog of BP/X bulges to date, which includes a category of candidate one-sided BP/X bulges. The catalog of one-sided structures is one of its kind sample, which likely contains samples of ongoing buckling events in edge-on galaxies.
    \item The fraction of one-sided BP/X bulges is always higher than the strong both-sided bulges at any given stellar mass. The fraction of both-sided BP/X bulges is always the highest for any stellar mass. At $\log (M_{*}/\Msun)=11.5$, the fraction of strong both-sided, one-sided and both-sided bulge is 18, 26, and 42 percent, respectively.
    \item We report a strong dependence of BP/X fraction on average stellar surface density of galaxy. The fraction of BP/X hosting galaxies increases with the surface density, reflecting bar formation in high surface density disk and subsequent BP/X structure growth.
    \item We use our catalog to investigate their role on stellar mass-size relation and stellar mass-stellar surface density relation. We found that galaxies with strong both-sided BP/X structures contribute to the scatter in these relations at high-mass end. However, the scaling relations of control and strong both-sided BP/X show significant overlap in their scatter.
\end{itemize}

This study provides a classification of BP/X bulges in a large sample of edge-on galaxies, along with their statistical properties and evolutionary trends. We demonstrate the effect of bar formation and evolution on the mass–size and mass–density scaling relations. The analysis uses single-component fitting for size measurements and empirical relations for dust correction and mass estimation, which introduce significant scatter in the scaling relations. Future studies employing detailed 2D decomposition for scale size measurements and SED fitting for stellar mass, star formation rate, dust content, and star formation history will be instrumental in the era of large-scale surveys, e.g., Euclid, LSST, 4MOST, and Roman. With the increase in spatial resolution, it will be possible to distinguish between classical bulge and weak BP/X bulge. We can find the real one-sided BP/X bulge galaxies from the class of one-sided galaxies. Additionally, our sample can be used to train AI models for studying large datasets.

\begin{acknowledgements}
We thank anonymous reviewer for providing constructive feedback and suggestion, which helped improve this paper. AK acknowledges support from ALMA fund with code 31220021 and from ANID BASAL project FB210003.

Funding for the Sloan Digital Sky Survey V has been provided by the Alfred P. Sloan Foundation, the Heising-Simons Foundation, the National Science Foundation, and the Participating Institutions. SDSS acknowledges support and resources from the Center for High-Performance Computing at the University of Utah. SDSS telescopes are located at Apache Point Observatory, funded by the Astrophysical Research Consortium and operated by New Mexico State University, and at Las Campanas Observatory, operated by the Carnegie Institution for Science. The SDSS web site is \url{www.sdss.org}.

SDSS is managed by the Astrophysical Research Consortium for the Participating Institutions of the SDSS Collaboration, including Caltech, The Carnegie Institution for Science, Chilean National Time Allocation Committee (CNTAC) ratified researchers, The Flatiron Institute, the Gotham Participation Group, Harvard University, Heidelberg University, The Johns Hopkins University, L'Ecole polytechnique f\'{e}d\'{e}rale de Lausanne (EPFL), Leibniz-Institut f\"{u}r Astrophysik Potsdam (AIP), Max-Planck-Institut f\"{u}r Astronomie (MPIA Heidelberg), Max-Planck-Institut f\"{u}r Extraterrestrische Physik (MPE), Nanjing University, National Astronomical Observatories of China (NAOC), New Mexico State University, The Ohio State University, Pennsylvania State University, Smithsonian Astrophysical Observatory, Space Telescope Science Institute (STScI), the Stellar Astrophysics Participation Group, Universidad Nacional Aut\'{o}noma de M\'{e}xico, University of Arizona, University of Colorado Boulder, University of Illinois at Urbana-Champaign, University of Toronto, University of Utah, University of Virginia, Yale University, and Yunnan University.

This publication uses data generated via the \href{https://www.zooniverse.org/}{zooniverse.org} platform, development of which is funded by generous support, including a Global Impact Award from Google, and by a grant from the Alfred P. Sloan Foundation.

We make use of Matplotlib \citep{matplotlib2007}, Numpy \citep{numpy2020}, Pandas \citep{pandas2010, pandas2020}, Scipy \citep{Scipy2020}. 
\end{acknowledgements}

\bibliographystyle{aa}
\bibliography{references.bib}

\begin{thebibliography}{128}
\expandafter\ifx\csname natexlab\endcsname\relax\def\natexlab#1{#1}\fi

\bibitem[{Aguerri {et~al.}(2009)Aguerri, M{\'e}ndez-Abreu, \& Corsini}]{aguerri2009population}
Aguerri, J., M{\'e}ndez-Abreu, J., \& Corsini, E. 2009, Astronomy \& Astrophysics, 495, 491

\bibitem[{Aihara {et~al.}(2011)Aihara, Prieto, An, Anderson, Aubourg, Balbinot, Beers, Berlind, Bickerton, Bizyaev, {et~al.}}]{aihara2011eighth}
Aihara, H., Prieto, C.~A., An, D., {et~al.} 2011, The Astrophysical Journal Supplement Series, 193, 29

\bibitem[{{Alonso} {et~al.}(2013){Alonso}, {Coldwell}, \& {Lambas}}]{Alonso.etal.2013}
{Alonso}, M.~S., {Coldwell}, G., \& {Lambas}, D.~G. 2013, \aap, 549, A141

\bibitem[{Athanassoula(2003)}]{athanassoula2003determines}
Athanassoula, E. 2003, Monthly Notices of the Royal Astronomical Society, 341, 1179

\bibitem[{Athanassoula(2005)}]{athanassoula2005nature}
Athanassoula, E. 2005, Monthly Notices of the Royal Astronomical Society, 358, 1477

\bibitem[{Athanassoula \& Martinez-Valpuesta(2008)}]{athanassoula2008boxy}
Athanassoula, E. \& Martinez-Valpuesta, I. 2008, in Chaos in Astronomy (Springer), 77--83

\bibitem[{Baba \& Kawata(2020)}]{baba2020age}
Baba, J. \& Kawata, D. 2020, Monthly Notices of the Royal Astronomical Society, 492, 4500

\bibitem[{{Barro} {et~al.}(2013){Barro}, {Faber}, {P{\'e}rez-Gonz{\'a}lez}, {Koo}, {Williams}, {Kocevski}, {Trump}, {Mozena}, {McGrath}, {van der Wel}, {Wuyts}, {Bell}, {Croton}, {Ceverino}, {Dekel}, {Ashby}, {Cheung}, {Ferguson}, {Fontana}, {Fang}, {Giavalisco}, {Grogin}, {Guo}, {Hathi}, {Hopkins}, {Huang}, {Koekemoer}, {Kartaltepe}, {Lee}, {Newman}, {Porter}, {Primack}, {Ryan}, {Rosario}, {Somerville}, {Salvato}, \& {Hsu}}]{Barro.etal.2013}
{Barro}, G., {Faber}, S.~M., {P{\'e}rez-Gonz{\'a}lez}, P.~G., {et~al.} 2013, \apj, 765, 104

\bibitem[{Bell {et~al.}(2003)Bell, McIntosh, Katz, \& Weinberg}]{bell2003optical}
Bell, E.~F., McIntosh, D.~H., Katz, N., \& Weinberg, M.~D. 2003, The Astrophysical Journal Supplement Series, 149, 289

\bibitem[{{Berentzen} {et~al.}(2007){Berentzen}, {Shlosman}, {Martinez-Valpuesta}, \& {Heller}}]{Berentzen.etal.2007}
{Berentzen}, I., {Shlosman}, I., {Martinez-Valpuesta}, I., \& {Heller}, C.~H. 2007, \apj, 666, 189

\bibitem[{{Bettoni} \& {Galletta}(1994)}]{Bettoni.Galletta.1994}
{Bettoni}, D. \& {Galletta}, G. 1994, \aap, 281, 1

\bibitem[{Blanton {et~al.}(2003)Blanton, Hogg, Bahcall, Brinkmann, Britton, Connolly, Csabai, Fukugita, Loveday, Meiksin, {et~al.}}]{blanton2003galaxy}
Blanton, M.~R., Hogg, D.~W., Bahcall, N.~A., {et~al.} 2003, The Astrophysical Journal, 592, 819

\bibitem[{{Boone} {et~al.}(2007){Boone}, {Baker}, {Schinnerer}, {Combes}, {Garc{\'\i}a-Burillo}, {Neri}, {Hunt}, {L{\'e}on}, {Krips}, {Tacconi}, \& {Eckart}}]{Boone.etal.2007}
{Boone}, F., {Baker}, A.~J., {Schinnerer}, E., {et~al.} 2007, \aap, 471, 113

\bibitem[{Bradley \& Developers(2023)}]{larry_bradley_2023_7859265}
Bradley, L. \& Developers, P. 2023, astropy/photutils: 1.6.0

\bibitem[{Bureau \& Athanassoula(2005)}]{bureau2005bar}
Bureau, M. \& Athanassoula, E. 2005, The Astrophysical Journal, 626, 159

\bibitem[{{Cervantes Sodi}(2017)}]{Sodi.2017}
{Cervantes Sodi}, B. 2017, \apj, 835, 80

\bibitem[{Ciambur \& Graham(2016)}]{ciambur2016quantifying}
Ciambur, B.~C. \& Graham, A.~W. 2016, Monthly Notices of the Royal Astronomical Society, 459, 1276

\bibitem[{Cole {et~al.}(2014)Cole, Debattista, Erwin, Earp, \& Ro{\v{s}}kar}]{cole2014formation}
Cole, D.~R., Debattista, V.~P., Erwin, P., Earp, S.~W., \& Ro{\v{s}}kar, R. 2014, Monthly Notices of the Royal Astronomical Society, 445, 3352

\bibitem[{{Collier}(2020)}]{Collier.2020}
{Collier}, A. 2020, \mnras, 492, 2241

\bibitem[{Combes {et~al.}(1990)Combes, Debbasch, Friedli, \& Pfenniger}]{combes1990box}
Combes, F., Debbasch, F., Friedli, D., \& Pfenniger, D. 1990, Astronomy and Astrophysics (ISSN 0004-6361), vol. 233, no. 1, July 1990, p. 82-95. Research supported by the Universite de Geneve and SNSF., 233, 82

\bibitem[{Combes \& Sanders(1981)}]{combes1981formation}
Combes, F. \& Sanders, R. 1981, Astronomy and Astrophysics, vol. 96, no. 1-2, Mar. 1981, p. 164-173., 96, 164

\bibitem[{{de S{\'a}-Freitas} {et~al.}(2023){de S{\'a}-Freitas}, {Fragkoudi}, {Gadotti}, {Falc{\'o}n-Barroso}, {Bittner}, {S{\'a}nchez-Bl{\'a}zquez}, {van de Ven}, {Bieri}, {Coccato}, {Coelho}, {Fahrion}, {Gon{\c{c}}alves}, {Kim}, {de Lorenzo-C{\'a}ceres}, {Martig}, {Mart{\'\i}n-Navarro}, {Mendez-Abreu}, {Neumann}, \& {Querejeta}}]{de_Sa-Freitas.etal.2023}
{de S{\'a}-Freitas}, C., {Fragkoudi}, F., {Gadotti}, D.~A., {et~al.} 2023, \aap, 671, A8

\bibitem[{De~Souza \& Dos~Anjos(1987)}]{de1987box}
De~Souza, R. \& Dos~Anjos, S. 1987, Astronomy and Astrophysics Supplement Series, 70, 465

\bibitem[{{de Vaucouleurs} {et~al.}(1991){de Vaucouleurs}, {de Vaucouleurs}, {Corwin}, {Buta}, {Paturel}, \& {Fouque}}]{de_Vaucouleurs.etal.1991}
{de Vaucouleurs}, G., {de Vaucouleurs}, A., {Corwin}, Herold~G., J., {et~al.} 1991, {Third Reference Catalogue of Bright Galaxies}

\bibitem[{{Debattista}(2018)}]{Debattista.2018}
{Debattista}, V. 2018, in The Galactic Bulge at the Crossroads, 10

\bibitem[{{Debattista} {et~al.}(2020){Debattista}, {Liddicott}, {Khachaturyants}, \& {Beraldo e Silva}}]{Debattista.etal.2020}
{Debattista}, V.~P., {Liddicott}, D.~J., {Khachaturyants}, T., \& {Beraldo e Silva}, L. 2020, \mnras, 498, 3334

\bibitem[{{Debattista} {et~al.}(2006){Debattista}, {Mayer}, {Carollo}, {Moore}, {Wadsley}, \& {Quinn}}]{Debattista.etal.2006}
{Debattista}, V.~P., {Mayer}, L., {Carollo}, C.~M., {et~al.} 2006, \apj, 645, 209

\bibitem[{{Dey} {et~al.}(2019){Dey}, {Schlegel}, {Lang}, {Blum}, {Burleigh}, {Fan}, {Findlay}, {Finkbeiner}, {Herrera}, {Juneau}, {Landriau}, {Levi}, {McGreer}, {Meisner}, {Myers}, {Moustakas}, {Nugent}, {Patej}, {Schlafly}, {Walker}, {Valdes}, {Weaver}, {Y{\`e}che}, {Zou}, {Zhou}, {Abareshi}, {Abbott}, {Abolfathi}, {Aguilera}, {Alam}, {Allen}, {Alvarez}, {Annis}, {Ansarinejad}, {Aubert}, {Beechert}, {Bell}, {BenZvi}, {Beutler}, {Bielby}, {Bolton}, {Brice{\~n}o}, {Buckley-Geer}, {Butler}, {Calamida}, {Carlberg}, {Carter}, {Casas}, {Castander}, {Choi}, {Comparat}, {Cukanovaite}, {Delubac}, {DeVries}, {Dey}, {Dhungana}, {Dickinson}, {Ding}, {Donaldson}, {Duan}, {Duckworth}, {Eftekharzadeh}, {Eisenstein}, {Etourneau}, {Fagrelius}, {Farihi}, {Fitzpatrick}, {Font-Ribera}, {Fulmer}, {G{\"a}nsicke}, {Gaztanaga}, {George}, {Gerdes}, {Gontcho}, {Gorgoni}, {Green}, {Guy}, {Harmer}, {Hernandez}, {Honscheid}, {Huang}, {James}, {Jannuzi}, {Jiang}, {Joyce}, {Karcher}, {Karkar}, {Kehoe}, {Kneib}, {Kueter-Young}, {Lan},
  {Lauer}, {Le Guillou}, {Le Van Suu}, {Lee}, {Lesser}, {Perreault Levasseur}, {Li}, {Mann}, {Marshall}, {Mart{\'\i}nez-V{\'a}zquez}, {Martini}, {du Mas des Bourboux}, {McManus}, {Meier}, {M{\'e}nard}, {Metcalfe}, {Mu{\~n}oz-Guti{\'e}rrez}, {Najita}, {Napier}, {Narayan}, {Newman}, {Nie}, {Nord}, {Norman}, {Olsen}, {Paat}, {Palanque-Delabrouille}, {Peng}, {Poppett}, {Poremba}, {Prakash}, {Rabinowitz}, {Raichoor}, {Rezaie}, {Robertson}, {Roe}, {Ross}, {Ross}, {Rudnick}, {Safonova}, {Saha}, {S{\'a}nchez}, {Savary}, {Schweiker}, {Scott}, {Seo}, {Shan}, {Silva}, {Slepian}, {Soto}, {Sprayberry}, {Staten}, {Stillman}, {Stupak}, {Summers}, {Sien Tie}, {Tirado}, {Vargas-Maga{\~n}a}, {Vivas}, {Wechsler}, {Williams}, {Yang}, {Yang}, {Yapici}, {Zaritsky}, {Zenteno}, {Zhang}, {Zhang}, {Zhou}, \& {Zhou}}]{Dey.etal.2019}
{Dey}, A., {Schlegel}, D.~J., {Lang}, D., {et~al.} 2019, \aj, 157, 168

\bibitem[{Di~Matteo {et~al.}(2019)Di~Matteo, Haywood, Lehnert, Katz, Khoperskov, Snaith, G{\'o}mez, \& Robichon}]{di2019milky}
Di~Matteo, P., Haywood, M., Lehnert, M., {et~al.} 2019, Astronomy \& Astrophysics, 632, A4

\bibitem[{{D{\'\i}az-Garc{\'\i}a} {et~al.}(2020){D{\'\i}az-Garc{\'\i}a}, {Moyano}, {Comer{\'o}n}, {Knapen}, {Salo}, \& {Bouquin}}]{Diaz-Garcia.etal.2020}
{D{\'\i}az-Garc{\'\i}a}, S., {Moyano}, F.~D., {Comer{\'o}n}, S., {et~al.} 2020, \aap, 644, A38

\bibitem[{D{\'\i}az-Garc{\'\i}a {et~al.}(2016)D{\'\i}az-Garc{\'\i}a, Salo, Laurikainen, \& Herrera-Endoqui}]{diaz2016characterization}
D{\'\i}az-Garc{\'\i}a, S., Salo, H., Laurikainen, E., \& Herrera-Endoqui, M. 2016, Astronomy \& Astrophysics, 587, A160

\bibitem[{Donohoe-Keyes {et~al.}(2019)Donohoe-Keyes, Martig, James, \& Kraljic}]{donohoe2019redistribution}
Donohoe-Keyes, C., Martig, M., James, P., \& Kraljic, K. 2019, Monthly Notices of the Royal Astronomical Society, 489, 4992

\bibitem[{Du \& McGaugh(2020)}]{du2020self}
Du, W. \& McGaugh, S.~S. 2020, The Astronomical Journal, 160, 122

\bibitem[{{Erwin}(2018)}]{2018MNRAS.474.5372E}
{Erwin}, P. 2018, \mnras, 474, 5372

\bibitem[{Erwin \& Debattista(2016)}]{erwin2016caught}
Erwin, P. \& Debattista, V.~P. 2016, The Astrophysical Journal Letters, 825, L30

\bibitem[{Erwin \& Debattista(2017)}]{erwin2017frequency}
Erwin, P. \& Debattista, V.~P. 2017, Monthly Notices of the Royal Astronomical Society, 468, 2058

\bibitem[{Fa{\'u}ndez-Abans \& de~Oliveira-Abans(1998)}]{faundez1998looking}
Fa{\'u}ndez-Abans, M. \& de~Oliveira-Abans, M. 1998, Astronomy and Astrophysics Supplement Series, 128, 289

\bibitem[{Fragkoudi {et~al.}(2017)Fragkoudi, Di~Matteo, Haywood, G{\'o}mez, Combes, Katz, \& Semelin}]{fragkoudi2017bars}
Fragkoudi, F., Di~Matteo, P., Haywood, M., {et~al.} 2017, Astronomy \& Astrophysics, 606, A47

\bibitem[{{Fraser-McKelvie} {et~al.}(2020){Fraser-McKelvie}, {Arag{\'o}n-Salamanca}, {Merrifield}, {Masters}, {Nair}, {Emsellem}, {Kraljic}, {Krishnarao}, {Andrews}, {Drory}, \& {Neumann}}]{Fraser-McKelvie.etal.2020}
{Fraser-McKelvie}, A., {Arag{\'o}n-Salamanca}, A., {Merrifield}, M., {et~al.} 2020, \mnras, 495, 4158

\bibitem[{Friedli \& Benz(1993)}]{friedli1993secular}
Friedli, D. \& Benz, W. 1993, Astronomy and Astrophysics (ISSN 0004-6361), vol. 268, no. 1, p. 65-85., 268, 65

\bibitem[{Friedli \& Benz(1995)}]{friedli1995secular}
Friedli, D. \& Benz, W. 1995, Astronomy and Astrophysics, v. 301, p. 649, 301, 649

\bibitem[{{Friedli} \& {Pfenniger}(1990)}]{Friedli.Pfenniger.1990}
{Friedli}, D. \& {Pfenniger}, D. 1990, in European Southern Observatory Conference and Workshop Proceedings, Vol.~35, European Southern Observatory Conference and Workshop Proceedings, ed. B.~J. {Jarvis} \& D.~M. {Terndrup}, 265

\bibitem[{{Galloway} {et~al.}(2015){Galloway}, {Willett}, {Fortson}, {Cardamone}, {Schawinski}, {Cheung}, {Lintott}, {Masters}, {Melvin}, \& {Simmons}}]{Galloway.etal.2015}
{Galloway}, M.~A., {Willett}, K.~W., {Fortson}, L.~F., {et~al.} 2015, \mnras, 448, 3442

\bibitem[{Garland {et~al.}(2024)Garland, Walmsley, Silcock, Potts, Smith, Simmons, Lintott, Smethurst, Dawson, Keel, Kruk, Mantha, Masters, O’Ryan, Popp, \& Thorne}]{10.1093/mnras/stae1620}
Garland, I.~L., Walmsley, M., Silcock, M.~S., {et~al.} 2024, Monthly Notices of the Royal Astronomical Society, 532, 2320

\bibitem[{{Ghosh} {et~al.}(2024){Ghosh}, {Fragkoudi}, {Di Matteo}, \& {Saha}}]{Ghosh.etal.2024}
{Ghosh}, S., {Fragkoudi}, F., {Di Matteo}, P., \& {Saha}, K. 2024, \aap, 683, A196

\bibitem[{{Ginsburg} {et~al.}(2019){Ginsburg}, {Sip{\H{o}}cz}, {Brasseur}, {Cowperthwaite}, {Craig}, {Deil}, {Guillochon}, {Guzman}, {Liedtke}, {Lian Lim}, {Lockhart}, {Mommert}, {Morris}, {Norman}, {Parikh}, {Persson}, {Robitaille}, {Segovia}, {Singer}, {Tollerud}, {de Val-Borro}, {Valtchanov}, {Woillez}, {Astroquery Collaboration}, \& {a subset of astropy Collaboration}}]{astroquery.Ginsburg.etal.2019}
{Ginsburg}, A., {Sip{\H{o}}cz}, B.~M., {Brasseur}, C.~E., {et~al.} 2019, \aj, 157, 98

\bibitem[{Harris {et~al.}(2020)Harris, Millman, van~der Walt, Gommers, Virtanen, Cournapeau, Wieser, Taylor, Berg, Smith, Kern, Picus, Hoyer, van Kerkwijk, Brett, Haldane, Fernández~del Río, Wiebe, Peterson, Gérard-Marchant, Sheppard, Reddy, Weckesser, Abbasi, Gohlke, \& Oliphant}]{numpy2020}
Harris, C.~R., Millman, K.~J., van~der Walt, S.~J., {et~al.} 2020, Nature, 585, 357–362

\bibitem[{Heller \& Shlosman(1994)}]{heller1994fueling}
Heller, C.~H. \& Shlosman, I. 1994, Astrophysical Journal, Part 1 (ISSN 0004-637X), vol. 424, no. 1, p. 84-105, 424, 84

\bibitem[{{Holmberg}(1958)}]{1958MeLuS.136....1H}
{Holmberg}, E. 1958, Meddelanden fran Lunds Astronomiska Observatorium Serie II, 136, 1

\bibitem[{{Hunter}(2007)}]{matplotlib2007}
{Hunter}, J.~D. 2007, Computing in Science Engineering, 9, 90

\bibitem[{Jarvis(1986)}]{jarvis1986search}
Jarvis, B.~J. 1986, The Astronomical Journal, 91, 65

\bibitem[{{Kalnajs}(1972)}]{Kalnajs.etal.1972}
{Kalnajs}, A.~J. 1972, \apj, 175, 63

\bibitem[{{Kataria} \& {Vivek}(2024)}]{Kataria.Vivek.2024}
{Kataria}, S.~K. \& {Vivek}, M. 2024, \mnras, 527, 3366

\bibitem[{Kautsch {et~al.}(2006)Kautsch, Grebel, Barazza, \& Gallagher}]{kautsch2006catalog}
Kautsch, S., Grebel, E., Barazza, F., \& Gallagher, J. 2006, Astronomy \& Astrophysics, 445, 765

\bibitem[{{Khoperskov} {et~al.}(2019){Khoperskov}, {Di Matteo}, {Gerhard}, {Katz}, {Haywood}, {Combes}, {Berczik}, \& {Gomez}}]{Khoperskov.etal.2019}
{Khoperskov}, S., {Di Matteo}, P., {Gerhard}, O., {et~al.} 2019, \aap, 622, L6

\bibitem[{Khoperskov {et~al.}(2018)Khoperskov, Haywood, Di~Matteo, Lehnert, \& Combes}]{khoperskov2018bar}
Khoperskov, S., Haywood, M., Di~Matteo, P., Lehnert, M., \& Combes, F. 2018, Astronomy \& Astrophysics, 609, A60

\bibitem[{Kim {et~al.}(2011)Kim, Saitoh, Jeon, Figer, Merritt, \& Wada}]{kim2011nuclear}
Kim, S.~S., Saitoh, T.~R., Jeon, M., {et~al.} 2011, The Astrophysical Journal Letters, 735, L11

\bibitem[{{Kruk} {et~al.}(2019){Kruk}, {Erwin}, {Debattista}, \& {Lintott}}]{Kruk.etal.2019}
{Kruk}, S.~J., {Erwin}, P., {Debattista}, V.~P., \& {Lintott}, C. 2019, \mnras, 490, 4721

\bibitem[{Kuijken \& Merrifield(1995)}]{kuijken1995establishing}
Kuijken, K. \& Merrifield, M.~R. 1995, arXiv preprint astro-ph/9501114

\bibitem[{{Kumar}(2023)}]{Kumar2023}
{Kumar}, A. 2023, arXiv e-prints, arXiv:2306.11045

\bibitem[{{Kumar} {et~al.}(2021){Kumar}, {Das}, \& {Kataria}}]{Kumar.etal.2021}
{Kumar}, A., {Das}, M., \& {Kataria}, S.~K. 2021, \mnras, 506, 98

\bibitem[{Kumar {et~al.}(2022)Kumar, Das, \& Kataria}]{kumar2022effect}
Kumar, A., Das, M., \& Kataria, S.~K. 2022, Monthly Notices of the Royal Astronomical Society, 509, 1262

\bibitem[{Kumar \& Kataria(2022)}]{kumar2022growth}
Kumar, A. \& Kataria, S.~K. 2022, Monthly Notices of the Royal Astronomical Society, 514, 2497

\bibitem[{{Lang} {et~al.}(2014){Lang}, {Holley-Bockelmann}, \& {Sinha}}]{Lang.etal.2014}
{Lang}, M., {Holley-Bockelmann}, K., \& {Sinha}, M. 2014, \apjl, 790, L33

\bibitem[{Laurikainen \& Salo(2016)}]{laurikainen2016observed}
Laurikainen, E. \& Salo, H. 2016, in Galactic Bulges (Springer), 77--106

\bibitem[{{Li} {et~al.}(2017){Li}, {Ho}, \& {Barth}}]{Li.etal.2017}
{Li}, Z.-Y., {Ho}, L.~C., \& {Barth}, A.~J. 2017, \apj, 845, 87

\bibitem[{Lintott {et~al.}(2010)Lintott, Schawinski, Bamford, Slosar, Land, Thomas, Edmondson, Masters, Nichol, Raddick, Szalay, Andreescu, Murray, \& Vandenberg}]{10.1111/j.1365-2966.2010.17432.x}
Lintott, C., Schawinski, K., Bamford, S., {et~al.} 2010, Monthly Notices of the Royal Astronomical Society, 410, 166

\bibitem[{Lintott {et~al.}(2008)Lintott, Schawinski, Slosar, Land, Bamford, Thomas, Raddick, Nichol, Szalay, Andreescu, {et~al.}}]{lintott2008galaxy}
Lintott, C.~J., Schawinski, K., Slosar, A., {et~al.} 2008, Monthly Notices of the Royal Astronomical Society, 389, 1179

\bibitem[{{{\L}okas}(2018)}]{Lokas.2018}
{{\L}okas}, E.~L. 2018, \apj, 857, 6

\bibitem[{{\L}okas(2019)}]{lokas2019anatomy}
{\L}okas, E.~L. 2019, Astronomy \& Astrophysics, 629, A52

\bibitem[{L{\"u}tticke {et~al.}(2000)L{\"u}tticke, Dettmar, \& Pohlen}]{lutticke2000box}
L{\"u}tticke, R., Dettmar, R.-J., \& Pohlen, M. 2000, Astronomy and Astrophysics Supplement Series, 145, 405

\bibitem[{Lynden-Bell(1979)}]{lynden1979mechanism}
Lynden-Bell, D. 1979, Monthly Notices of the Royal Astronomical Society, 187, 101

\bibitem[{{Marchuk} {et~al.}(2022){Marchuk}, {Smirnov}, {Sotnikova}, {Bunakalya}, {Savchenko}, {Reshetnikov}, {Usachev}, {Tikhonenko}, {Zozulia}, \& {Zakharova}}]{Marchuk.etal.2022}
{Marchuk}, A.~A., {Smirnov}, A.~A., {Sotnikova}, N.~Y., {et~al.} 2022, \mnras, 512, 1371

\bibitem[{Martin \& Friedli(1997)}]{martin1997star}
Martin, P. \& Friedli, D. 1997, Astronomy and Astrophysics, v. 326, p. 449-464, 326, 449

\bibitem[{{Martinez-Valpuesta} {et~al.}(2006){Martinez-Valpuesta}, {Shlosman}, \& {Heller}}]{Martinez-Valpuesta.etal.2006}
{Martinez-Valpuesta}, I., {Shlosman}, I., \& {Heller}, C. 2006, \apj, 637, 214

\bibitem[{Masters {et~al.}(2011)Masters, Nichol, Hoyle, Lintott, Bamford, Edmondson, Fortson, Keel, Schawinski, Smith, {et~al.}}]{masters2011galaxy}
Masters, K.~L., Nichol, R.~C., Hoyle, B., {et~al.} 2011, Monthly Notices of the Royal Astronomical Society, 411, 2026

\bibitem[{{Melvin} {et~al.}(2014){Melvin}, {Masters}, {Lintott}, {Nichol}, {Simmons}, {Bamford}, {Casteels}, {Cheung}, {Edmondson}, {Fortson}, {Schawinski}, {Skibba}, {Smith}, \& {Willett}}]{Melvin.etal.2014}
{Melvin}, T., {Masters}, K., {Lintott}, C., {et~al.} 2014, \mnras, 438, 2882

\bibitem[{Merritt \& Sellwood(1994)}]{merritt1994bending}
Merritt, D. \& Sellwood, J. 1994, Astrophysical Journal, Part 1 (ISSN 0004-637X), vol. 425, no. 2, p. 551-567, 425, 551

\bibitem[{{Minchev} {et~al.}(2011){Minchev}, {Famaey}, {Combes}, {Di Matteo}, {Mouhcine}, \& {Wozniak}}]{Minchev.etal.2011}
{Minchev}, I., {Famaey}, B., {Combes}, F., {et~al.} 2011, \aap, 527, A147

\bibitem[{{Miwa} \& {Noguchi}(1998)}]{Miwa.Noguchi.1998}
{Miwa}, T. \& {Noguchi}, M. 1998, \apj, 499, 149

\bibitem[{{Naab} {et~al.}(2009){Naab}, {Johansson}, \& {Ostriker}}]{Naab.etal.2009}
{Naab}, T., {Johansson}, P.~H., \& {Ostriker}, J.~P. 2009, \apjl, 699, L178

\bibitem[{{Nair} \& {Abraham}(2010)}]{Nair.Abraham.2010}
{Nair}, P.~B. \& {Abraham}, R.~G. 2010, \apjl, 714, L260

\bibitem[{{Nelson} {et~al.}(2016){Nelson}, {van Dokkum}, {F{\"o}rster Schreiber}, {Franx}, {Brammer}, {Momcheva}, {Wuyts}, {Whitaker}, {Skelton}, {Fumagalli}, {Hayward}, {Kriek}, {Labb{\'e}}, {Leja}, {Rix}, {Tacconi}, {van der Wel}, {van den Bosch}, {Oesch}, {Dickey}, \& {Ulf Lange}}]{Nelson.etal.2016}
{Nelson}, E.~J., {van Dokkum}, P.~G., {F{\"o}rster Schreiber}, N.~M., {et~al.} 2016, \apj, 828, 27

\bibitem[{{Ness} \& {Lang}(2016)}]{Ness.etal.2016}
{Ness}, M. \& {Lang}, D. 2016, \aj, 152, 14

\bibitem[{{Ostriker} \& {Peebles}(1973)}]{Ostriker.Peebles.1973}
{Ostriker}, J.~P. \& {Peebles}, P.~J.~E. 1973, \apj, 186, 467

\bibitem[{pandas~development team(2020)}]{pandas2020}
pandas~development team, T. 2020, pandas-dev/pandas: Pandas

\bibitem[{{Peng} {et~al.}(2010){Peng}, {Ho}, {Impey}, \& {Rix}}]{Peng.etal.2010}
{Peng}, C.~Y., {Ho}, L.~C., {Impey}, C.~D., \& {Rix}, H.-W. 2010, \aj, 139, 2097

\bibitem[{Pfenniger \& Friedli(1991)}]{pfenniger1991structure}
Pfenniger, D. \& Friedli, D. 1991, Astronomy and Astrophysics (ISSN 0004-6361), vol. 252, no. 1, Dec. 1991, p. 75-93. Research supported by Observatoire de Geneve and SNSF., 252, 75

\bibitem[{Polyachenko \& Polyachenko(2003)}]{polyachenko2003unified}
Polyachenko, V. \& Polyachenko, E. 2003, Astronomy Letters, 29, 447

\bibitem[{Quillen(2002)}]{quillen2002growth}
Quillen, A.~C. 2002, The Astronomical Journal, 124, 722

\bibitem[{Quillen {et~al.}(2014)Quillen, Minchev, Sharma, Qin, \& Di~Matteo}]{quillen2014vertical}
Quillen, A.~C., Minchev, I., Sharma, S., Qin, Y.-J., \& Di~Matteo, P. 2014, Monthly Notices of the Royal Astronomical Society, 437, 1284

\bibitem[{Raha {et~al.}(1991)Raha, Sellwood, James, \& Kahn}]{raha1991dynamical}
Raha, N., Sellwood, J., James, R., \& Kahn, F. 1991, nature, 352, 411

\bibitem[{Saha {et~al.}(2013)Saha, Pfenniger, \& Taam}]{saha2013meridional}
Saha, K., Pfenniger, D., \& Taam, R.~E. 2013, The Astrophysical Journal, 764, 123

\bibitem[{Savchenko {et~al.}(2017)Savchenko, Sotnikova, Mosenkov, Reshetnikov, \& Bizyaev}]{savchenko2017measuring}
Savchenko, S., Sotnikova, N.~Y., Mosenkov, A., Reshetnikov, V., \& Bizyaev, D. 2017, Monthly Notices of the Royal Astronomical Society, 471, 3261

\bibitem[{{Savchenko} {et~al.}(2023){Savchenko}, {Poliakov}, {Mosenkov}, {Smirnov}, {Marchuk}, {Il'in}, {Gontcharov}, {Seguine}, \& {Baes}}]{2023MNRAS.524.4729S}
{Savchenko}, S.~S., {Poliakov}, D.~M., {Mosenkov}, A.~V., {et~al.} 2023, \mnras, 524, 4729

\bibitem[{Sellwood \& Gerhard(2020)}]{sellwood2020three}
Sellwood, J. \& Gerhard, O. 2020, Monthly Notices of the Royal Astronomical Society, 495, 3175

\bibitem[{{Sellwood}(2014)}]{Sellwood2014}
{Sellwood}, J.~A. 2014, Reviews of Modern Physics, 86, 1

\bibitem[{Seo {et~al.}(2019)Seo, Kim, Kwak, Hsieh, Han, \& Hopkins}]{seo2019effects}
Seo, W.-Y., Kim, W.-T., Kwak, S., {et~al.} 2019, The Astrophysical Journal, 872, 5

\bibitem[{Shao {et~al.}(2007)Shao, Xiao, Shen, Mo, Xia, \& Deng}]{shao2007inclination}
Shao, Z., Xiao, Q., Shen, S., {et~al.} 2007, The Astrophysical Journal, 659, 1159

\bibitem[{Shaw(1987)}]{shaw1987nature}
Shaw, M.~A. 1987, Monthly Notices of the Royal Astronomical Society, 229, 691

\bibitem[{{Shen} {et~al.}(2003){Shen}, {Mo}, {White}, {Blanton}, {Kauffmann}, {Voges}, {Brinkmann}, \& {Csabai}}]{Shen.etal.2003}
{Shen}, S., {Mo}, H.~J., {White}, S. D.~M., {et~al.} 2003, \mnras, 343, 978

\bibitem[{{Shlosman} {et~al.}(1989){Shlosman}, {Frank}, \& {Begelman}}]{Shlosman.etal.1989}
{Shlosman}, I., {Frank}, J., \& {Begelman}, M.~C. 1989, \nat, 338, 45

\bibitem[{{Silva-Lima} {et~al.}(2022){Silva-Lima}, {Martins}, {Coelho}, \& {Gadotti}}]{Silva-Lima.etal.2022}
{Silva-Lima}, L.~A., {Martins}, L.~P., {Coelho}, P. R.~T., \& {Gadotti}, D.~A. 2022, \aap, 661, A105

\bibitem[{Simard {et~al.}(2011)Simard, Mendel, Patton, Ellison, \& McConnachie}]{simard2011catalog}
Simard, L., Mendel, J.~T., Patton, D.~R., Ellison, S.~L., \& McConnachie, A.~W. 2011, The Astrophysical Journal Supplement Series, 196, 11

\bibitem[{Smirnov \& Sotnikova(2018)}]{smirnov2018determines}
Smirnov, A.~A. \& Sotnikova, N.~Y. 2018, Monthly Notices of the Royal Astronomical Society, 481, 4058

\bibitem[{{Smirnov} \& {Sotnikova}(2019)}]{Smirnov.etal.2019}
{Smirnov}, A.~A. \& {Sotnikova}, N.~Y. 2019, \mnras, 485, 1900

\bibitem[{Sormani {et~al.}(2022)Sormani, Gerhard, Portail, Vasiliev, \& Clarke}]{sormani2022stellar}
Sormani, M.~C., Gerhard, O., Portail, M., Vasiliev, E., \& Clarke, J. 2022, Monthly Notices of the Royal Astronomical Society: Letters, 514, L1

\bibitem[{Spinoso {et~al.}(2016)Spinoso, Bonoli, Dotti, Mayer, Madau, \& Bellovary}]{spinoso2016bar}
Spinoso, D., Bonoli, S., Dotti, M., {et~al.} 2016, Monthly Notices of the Royal Astronomical Society, 465, 3729

\bibitem[{Stoughton {et~al.}(2002)Stoughton, Lupton, Bernardi, Blanton, Burles, Castander, Connolly, Eisenstein, Frieman, Hennessy, {et~al.}}]{stoughton2002sloan}
Stoughton, C., Lupton, R.~H., Bernardi, M., {et~al.} 2002, The Astronomical Journal, 123, 485

\bibitem[{{Toomre}(1981)}]{Toomre.etal.1981}
{Toomre}, A. 1981, in Structure and Evolution of Normal Galaxies, ed. S.~M. {Fall} \& D.~{Lynden-Bell}, 111--136

\bibitem[{{Trujillo} {et~al.}(2006){Trujillo}, {F{\"o}rster Schreiber}, {Rudnick}, {Barden}, {Franx}, {Rix}, {Caldwell}, {McIntosh}, {Toft}, {H{\"a}ussler}, {Zirm}, {van Dokkum}, {Labb{\'e}}, {Moorwood}, {R{\"o}ttgering}, {van der Wel}, {van der Werf}, \& {van Starkenburg}}]{Trujillo.etal.2006}
{Trujillo}, I., {F{\"o}rster Schreiber}, N.~M., {Rudnick}, G., {et~al.} 2006, \apj, 650, 18

\bibitem[{{Tully} \& {Fisher}(1977)}]{1977A&A....54..661T}
{Tully}, R.~B. \& {Fisher}, J.~R. 1977, \aap, 54, 661

\bibitem[{Van~der Kruit \& Searle(1981)}]{van1981surface}
Van~der Kruit, P. \& Searle, L. 1981, Astronomy and Astrophysics, vol. 95, no. 1, Feb. 1981, p. 105-115., 95, 105

\bibitem[{{van der Kruit} \& {Freeman}(2011)}]{Kruit.Freeman.2011}
{van der Kruit}, P.~C. \& {Freeman}, K.~C. 2011, \araa, 49, 301

\bibitem[{{van der Wel} {et~al.}(2014){van der Wel}, {Franx}, {van Dokkum}, {Skelton}, {Momcheva}, {Whitaker}, {Brammer}, {Bell}, {Rix}, {Wuyts}, {Ferguson}, {Holden}, {Barro}, {Koekemoer}, {Chang}, {McGrath}, {H{\"a}ussler}, {Dekel}, {Behroozi}, {Fumagalli}, {Leja}, {Lundgren}, {Maseda}, {Nelson}, {Wake}, {Patel}, {Labb{\'e}}, {Faber}, {Grogin}, \& {Kocevski}}]{van_der_Wel.etal.2014}
{van der Wel}, A., {Franx}, M., {van Dokkum}, P.~G., {et~al.} 2014, \apj, 788, 28

\bibitem[{{van Dokkum} {et~al.}(2010){van Dokkum}, {Whitaker}, {Brammer}, {Franx}, {Kriek}, {Labb{\'e}}, {Marchesini}, {Quadri}, {Bezanson}, {Illingworth}, {Muzzin}, {Rudnick}, {Tal}, \& {Wake}}]{van_Dokkum.etal.2010}
{van Dokkum}, P.~G., {Whitaker}, K.~E., {Brammer}, G., {et~al.} 2010, \apj, 709, 1018

\bibitem[{{Veilleux} {et~al.}(1999){Veilleux}, {Bland-Hawthorn}, \& {Cecil}}]{Veilleux.etal.1999}
{Veilleux}, S., {Bland-Hawthorn}, J., \& {Cecil}, G. 1999, \aj, 118, 2108

\bibitem[{Virtanen {et~al.}(2020{\natexlab{a}})Virtanen, Gommers, Oliphant, Haberland, Reddy, Cournapeau, Burovski, Peterson, Weckesser, Bright, {van der Walt}, Brett, Wilson, Millman, Mayorov, Nelson, Jones, Kern, Larson, Carey, Polat, Feng, Moore, {VanderPlas}, Laxalde, Perktold, Cimrman, Henriksen, Quintero, Harris, Archibald, Ribeiro, Pedregosa, {van Mulbregt}, \& {SciPy 1.0 Contributors}}]{2020SciPy-NMeth}
Virtanen, P., Gommers, R., Oliphant, T.~E., {et~al.} 2020{\natexlab{a}}, Nature Methods, 17, 261

\bibitem[{Virtanen {et~al.}(2020{\natexlab{b}})Virtanen, Gommers, Oliphant, Haberland, Reddy, Cournapeau, Burovski, Peterson, Weckesser, Bright, {van der Walt}, Brett, Wilson, Millman, Mayorov, Nelson, Jones, Kern, Larson, Carey, Polat, Feng, Moore, {VanderPlas}, Laxalde, Perktold, Cimrman, Henriksen, Quintero, Harris, Archibald, Ribeiro, Pedregosa, {van Mulbregt}, \& {SciPy 1.0 Contributors}}]{Scipy2020}
Virtanen, P., Gommers, R., Oliphant, T.~E., {et~al.} 2020{\natexlab{b}}, Nature Methods, 17, 261

\bibitem[{{W}es {M}c{K}inney(2010)}]{pandas2010}
{W}es {M}c{K}inney. 2010, in {P}roceedings of the 9th {P}ython in {S}cience {C}onference, ed. {S}t\'efan van~der {W}alt \& {J}arrod {M}illman, 56 -- 61

\bibitem[{Wozniak(2007)}]{wozniak2007distribution}
Wozniak, H. 2007, Astronomy \& Astrophysics, 465, L1

\bibitem[{{Xiang} {et~al.}(2021){Xiang}, {Nataf}, {Athanassoula}, {Zakamska}, {Rowlands}, {Masters}, {Fraser-McKelvie}, {Drory}, \& {Kraljic}}]{Xiang.etal.2021}
{Xiang}, K.~M., {Nataf}, D.~M., {Athanassoula}, E., {et~al.} 2021, \apj, 909, 125

\bibitem[{{York} {et~al.}(2000){York}, {Adelman}, {Anderson}, {Anderson}, {Annis}, {Bahcall}, {Bakken}, {Barkhouser}, {Bastian}, {Berman}, {Boroski}, {Bracker}, {Briegel}, {Briggs}, {Brinkmann}, {Brunner}, {Burles}, {Carey}, {Carr}, {Castander}, {Chen}, {Colestock}, {Connolly}, {Crocker}, {Csabai}, {Czarapata}, {Davis}, {Doi}, {Dombeck}, {Eisenstein}, {Ellman}, {Elms}, {Evans}, {Fan}, {Federwitz}, {Fiscelli}, {Friedman}, {Frieman}, {Fukugita}, {Gillespie}, {Gunn}, {Gurbani}, {de Haas}, {Haldeman}, {Harris}, {Hayes}, {Heckman}, {Hennessy}, {Hindsley}, {Holm}, {Holmgren}, {Huang}, {Hull}, {Husby}, {Ichikawa}, {Ichikawa}, {Ivezi{\'c}}, {Kent}, {Kim}, {Kinney}, {Klaene}, {Kleinman}, {Kleinman}, {Knapp}, {Korienek}, {Kron}, {Kunszt}, {Lamb}, {Lee}, {Leger}, {Limmongkol}, {Lindenmeyer}, {Long}, {Loomis}, {Loveday}, {Lucinio}, {Lupton}, {MacKinnon}, {Mannery}, {Mantsch}, {Margon}, {McGehee}, {McKay}, {Meiksin}, {Merelli}, {Monet}, {Munn}, {Narayanan}, {Nash}, {Neilsen}, {Neswold}, {Newberg}, {Nichol}, {Nicinski},
  {Nonino}, {Okada}, {Okamura}, {Ostriker}, {Owen}, {Pauls}, {Peoples}, {Peterson}, {Petravick}, {Pier}, {Pope}, {Pordes}, {Prosapio}, {Rechenmacher}, {Quinn}, {Richards}, {Richmond}, {Rivetta}, {Rockosi}, {Ruthmansdorfer}, {Sandford}, {Schlegel}, {Schneider}, {Sekiguchi}, {Sergey}, {Shimasaku}, {Siegmund}, {Smee}, {Smith}, {Snedden}, {Stone}, {Stoughton}, {Strauss}, {Stubbs}, {SubbaRao}, {Szalay}, {Szapudi}, {Szokoly}, {Thakar}, {Tremonti}, {Tucker}, {Uomoto}, {Vanden Berk}, {Vogeley}, {Waddell}, {Wang}, {Watanabe}, {Weinberg}, {Yanny}, {Yasuda}, \& {SDSS Collaboration}}]{York.etal.2000}
{York}, D.~G., {Adelman}, J., {Anderson}, Jr., J.~E., {et~al.} 2000, \aj, 120, 1579

\bibitem[{Yoshino \& Yamauchi(2015)}]{yoshino2015box}
Yoshino, A. \& Yamauchi, C. 2015, Monthly Notices of the Royal Astronomical Society, 446, 3749

\bibitem[{Yuan {et~al.}(2013)Yuan, Liu, \& Xiang}]{yuan2013empirical}
Yuan, H.-B., Liu, X.-W., \& Xiang, M.-S. 2013, Monthly Notices of the Royal Astronomical Society, 430, 2188

\bibitem[{Zozulia {et~al.}(2024{\natexlab{a}})Zozulia, Smirnov, \& Sotnikova}]{10.1093/mnras/stae702}
Zozulia, V.~D., Smirnov, A.~A., \& Sotnikova, N.~Y. 2024{\natexlab{a}}, Monthly Notices of the Royal Astronomical Society, 529, 4405

\bibitem[{Zozulia {et~al.}(2024{\natexlab{b}})Zozulia, Smirnov, Sotnikova, \& Marchuk}]{zozulia2024boxy}
Zozulia, V.~D., Smirnov, A.~A., Sotnikova, N.~Y., \& Marchuk, A.~A. 2024{\natexlab{b}}, Astronomy \& Astrophysics, 692, A145

\bibitem[{Zozulia {et~al.}(2025)Zozulia, Sotnikova, \& Smirnov}]{zozulia2025phase}
Zozulia, V.~D., Sotnikova, N.~Y., \& Smirnov, A.~A. 2025, Astronomy \& Astrophysics, 698, L26

\end{thebibliography}

\newpage
\begin{appendix}

\section{More examples of sample galaxies}
\label{app:more_bpx_example}
\begin{figure}[!h]
    \centering
    \includegraphics[width=\linewidth]{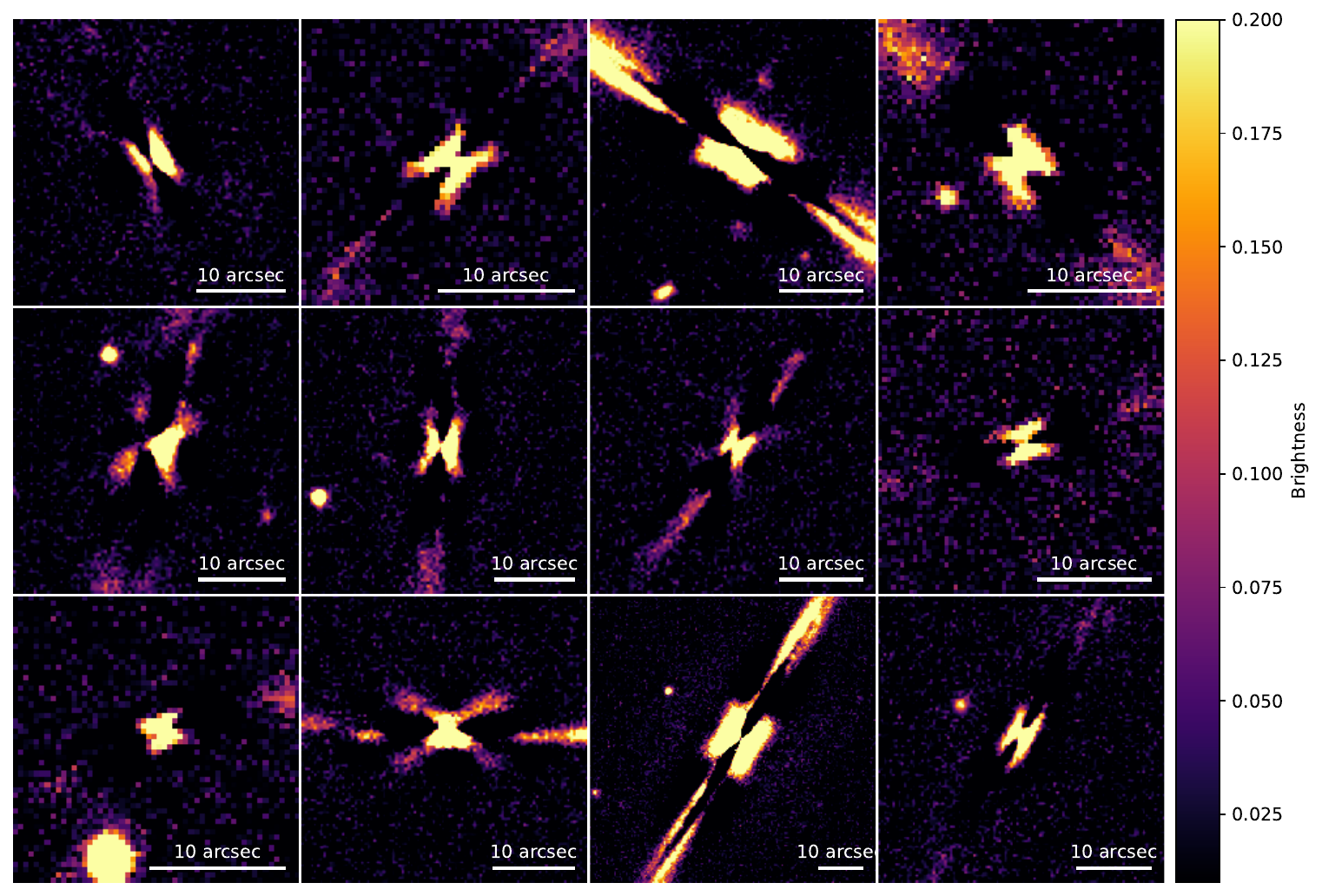}
    \caption{Strong both-sided residual images where the X shape can be clearly distinguished from the background.}
    \label{fig:more_strong_bpx}
\end{figure}

\begin{figure}[!h]
    \centering
    \includegraphics[width=\linewidth]{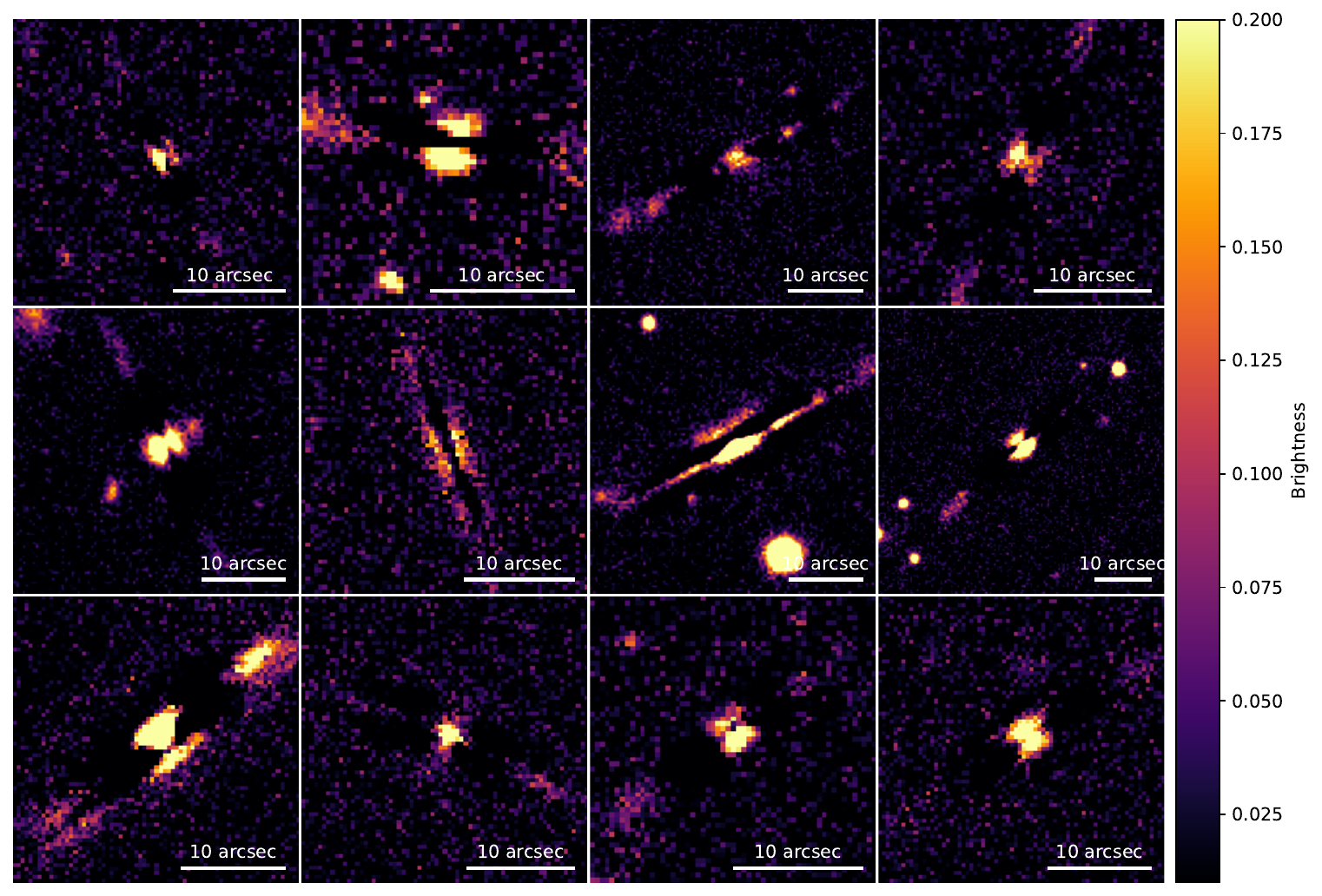}
    \caption{The residual images classified to have weak both-sided bulge due the presence of some nonzero residue at the center of the galaxy but the same is not clearly an X shape. }
    \label{fig:more_weak_bpx}
\end{figure}

\begin{figure}[!h]
    \centering
    \includegraphics[width=\linewidth]{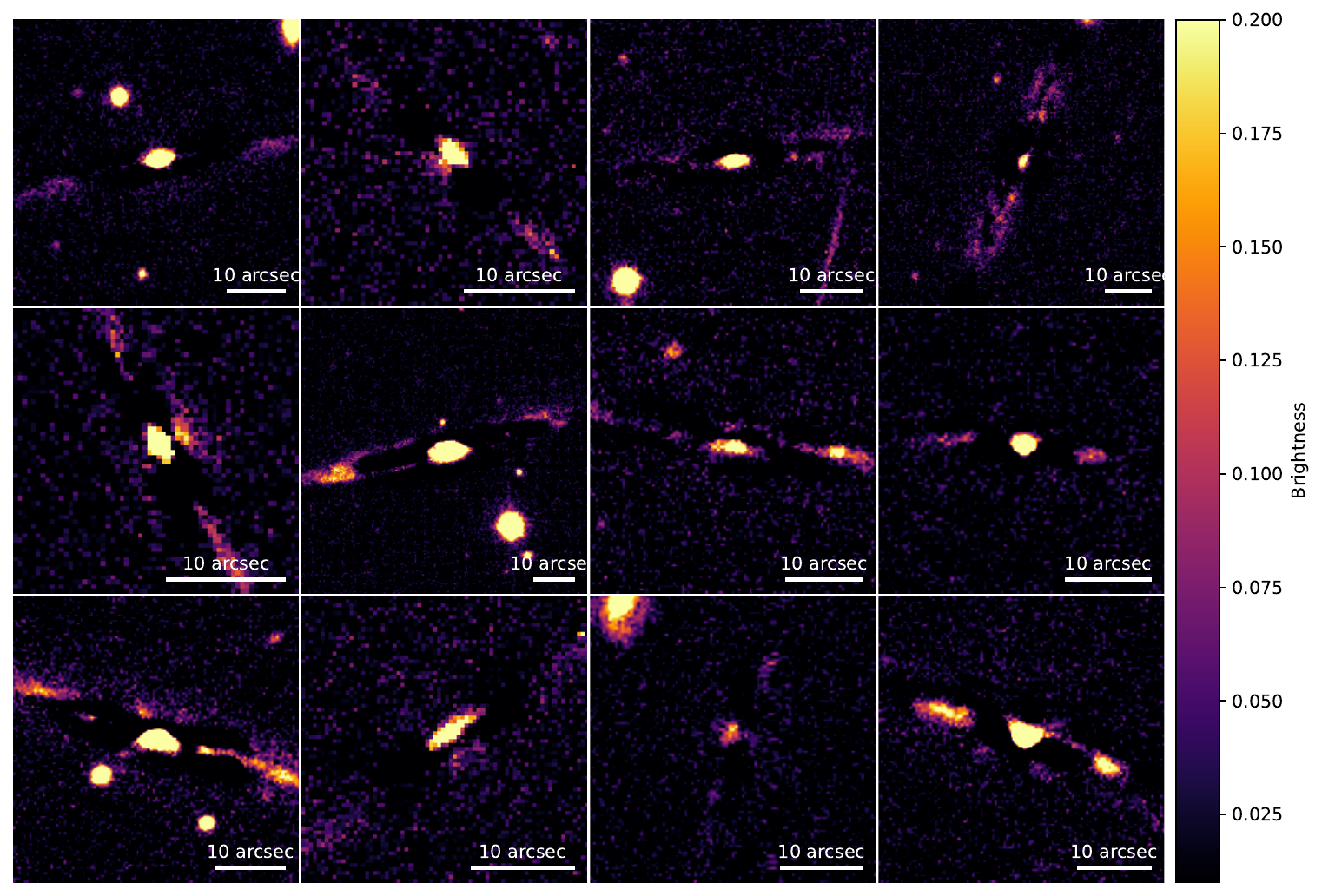}
    \caption{The residual images classified to have one-sided bulge due the presence of some nonzero residue at one side of the mid-plane of galaxy.}
    \label{fig:more_one_sided}
\end{figure}

In order to provide a clear overview of the different categories of galaxies with BP/X bulges defined in this study, we present mosaic images for each category. Fig.~\ref{fig:more_strong_bpx} shows a mosaic of strong, both-sided BP/X bulges; Fig.~\ref{fig:more_weak_bpx} displays a mosaic of weak, both-sided BP/X bulges; and Fig.~\ref{fig:more_one_sided} exhibits a mosaic of one-sided BP/X bulges. Each panel in these images shows the residuals obtained by subtracting the model images from the observed ones. The color scales in each panel are adjusted to enhance the visibility of structural features.

\section{Effect of measuring surface density using scale radius}
\label{app:eff_of_scale_radius}
\begin{figure}[!h]
    \centering
    \includegraphics[width=\linewidth]{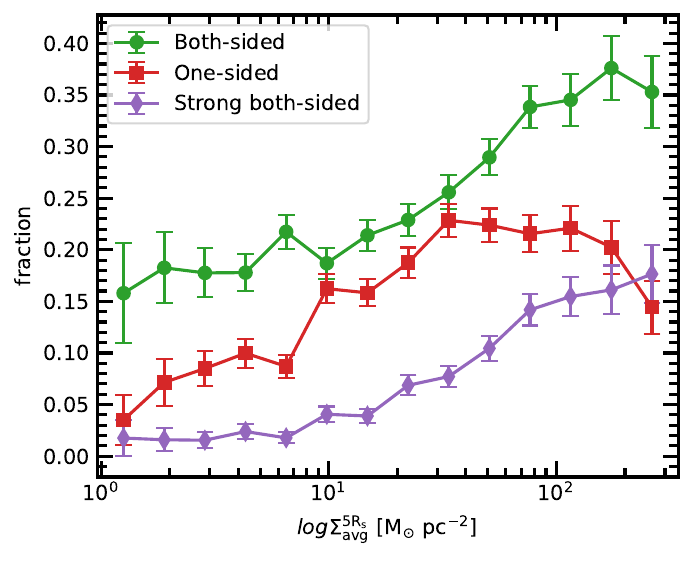}
    \caption{The fraction of both-sided (green), one-sided (red), and strong both-sided (purple) BP/X structures as a function of average surface density of galaxies. The high surface density disks host more BP/X structures than the low-surface density disks using scale radius.}
    \label{app:bpx_fraction_dens}
\end{figure}

In Section~\ref{sec:stellar_surface_density}, we discussed the dependence of the BP/X bulge-hosting galaxy fraction on the average stellar surface density, measured by assuming all stellar mass lies within the Petrosian radius. Here, we examine the effect of using the disk scale radius instead for calculating the average stellar density. Specifically, we compute the average stellar surface density under the assumption that all stellar mass is contained within five scale radii, and define it as $\rm \Sigma^{5R_{s}}_{avg} = M_{*}/(25\pi R_{s}^{2})$. Using a larger or smaller number of scale radii for this calculation does not affect the analysis qualitatively, as it merely scales the density by a constant factor. In Fig.~\ref{app:bpx_fraction_dens}, we present the BP/X fraction as a function of $\rm \Sigma^{5R_{s}}_{avg}$ for three bulge categories. For comparison with Fig.~\ref{fig:bpx_frac_with_dens}, we limit the average density range in the range of that shown in Fig.~\ref{fig:bpx_frac_with_dens}, scaled by a factor of $\approx 0.3$, which corresponds to the average value of $(R_{p})^{2}/(5R_{s})^{2}$. The results are highly consistent with those presented in Section~\ref{sec:stellar_surface_density}, showing that the fraction of BP/X bulges increases with increasing stellar surface density. There are very small differences at low and high surface density ends likely due to small sample size.

\begin{figure}[!h]
    \centering
    \includegraphics[width=\linewidth]{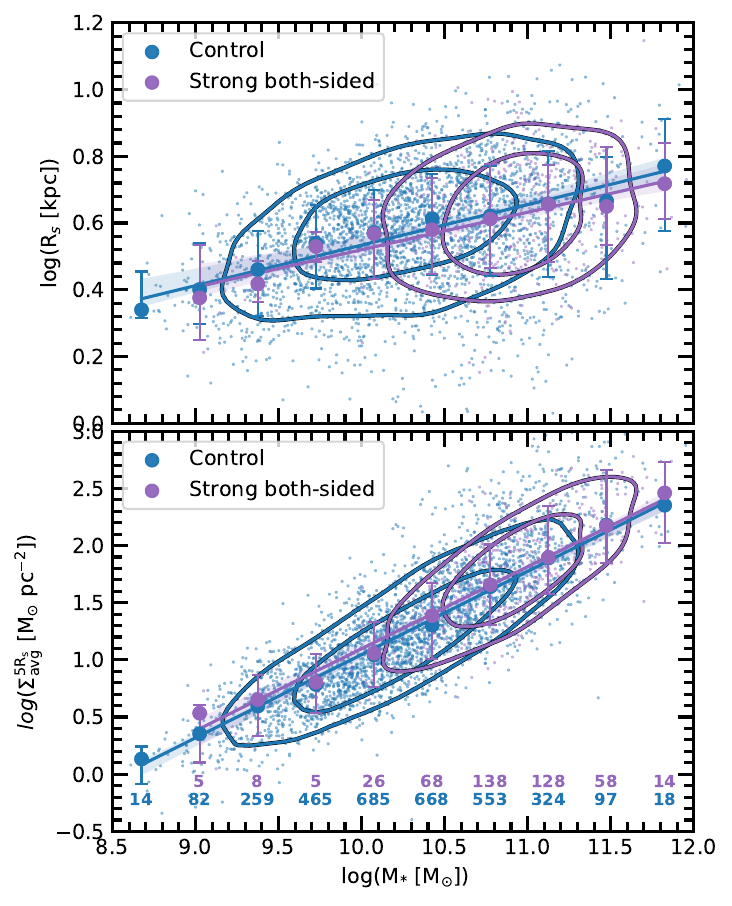}
    \caption{The stellar mass–size relation (top panel) and stellar mass–stellar surface density relation (bottom panel) for galaxies hosting BP/X structures. The control and strong both-sided samples are displayed in blue and purple, respectively. Both samples now exhibit nearly similar trends with negligible differences.}
    \label{app:scaling_relation}
\end{figure}

Similar to the mass–density relation discussed in Section~\ref{sec:effect_of_bpx_on_galaxy_evolution}, which assumes total mass within the Petrosian radius, we investigate the effect of using the scale radius in Fig.~\ref{app:scaling_relation}. Here, we show the mass–size relation in the top panel and the mass–density relation in the bottom panel, using the scale radius instead of the Petrosian radius. In contrast to the trends in Fig.~\ref{fig:log-relation}, both the control and strong both-sided BP/X bulge samples now show similar trends. This is likely due to the effects of bar thickening and dust distribution in galaxies, which influence scale radius measurements.

\end{appendix}

\end{document}